\documentclass[aip,jmp,reprint,amsmath,amssymb,onecolumn]{revtex4-1}

\usepackage{amsthm,bm,graphicx}

\theoremstyle{plain}\newtheorem{thm}{Theorem}
\theoremstyle{remark}\newtheorem*{rem}{Remark}

\newcommand\gS{\text{\normalfont S}}
\newcommand\gGL{\text{\normalfont GL}}
\newcommand\gSL{\text{\normalfont SL}}
\newcommand\gU{\text{\normalfont U}}
\newcommand\gO{\text{\normalfont O}}
\newcommand\gSO{\text{\normalfont SO}}
\newcommand\gPin{\text{\normalfont Pin}}
\newcommand\gSp{\text{\normalfont Sp}}
\newcommand\gMp{\text{\normalfont Mp}}

\newcommand\lA{\mathfrak{A}}
\newcommand\lh{\mathfrak{h}}
\newcommand\lgl{\mathfrak{gl}}
\newcommand\lsl{\mathfrak{sl}}
\newcommand\lo{\mathfrak{o}}
\newcommand\lsp{\mathfrak{sp}}
\newcommand\lsu{\mathfrak{su}}
\newcommand\lb{\mathfrak{b}}

\begin{document}

\title{Fock space dualities}

\author{K. Neerg\aa rd}

\affiliation{Fjordtoften 17, 4700 N\ae stved, Denmark}

\begin{abstract}

A general theorem due to Howe of dual action of a classical group and
a certain non-associative algebra on a space of symmetric or
alternating tensors is reformulated in a setting of second
quantization, and familiar examples in atomic and nuclear physics are
discussed. The special case of orthogonal-orthogonal duality is
treated in detail. It is shown that, like it was done by Helmers more
than half a century ago in the analogous case of symplectic-symplectic
duality, one can base a proof of the orthogonal-orthogonal duality
theorem and a precise characterization of the relation between the
equivalence classes of the dually related irreducible representations
on a calculation of characters by combining it, in this case, with an
analysis of the representation of a reflection. Young diagrams for the
description of equivalence classes of irreducible representations of
orthogonal Lie algebras are introduced. The properties of a reflection
of the number non-conserving part in the dual relationship between
orthogonal Lie algebras corroborate a picture of an almost perfect
symmetry between the partners.

\end{abstract}

\maketitle

\section{\label{sec:intr}Introduction}

This article is a sequel of Ref.~\onlinecite{ref:Nee19} with multiple
aims. One aim is to relate my results in Ref.~\onlinecite{ref:Nee19},
and related results in the literature, to a very general duality
theorem due to Howe. This theorem was proved in
Ref.~\onlinecite{ref:How89}, which I happened to read only after
Ref.~\onlinecite{ref:Nee19} was published. It has many special cases
which were known to and applied by physicists before
Ref.~\onlinecite{ref:How89}, but Howe's theorem places all of this in
a nice, unifying picture, as I intend to demonstrate in
Sec.~\ref{sec:exampl}. Howe's article is written in a professional,
mathematical language, which may feel foreign to workers in atomic and
nuclear spectrocopy. This may be one reason why his result, while
appearing in preprint in 1976, seems virtually unknown to this sector
of the physics community. My first task, before entering the
discussion of physical applications, is therefore to reformulate
Howe's duality theorem in a, to this readership, hopefully more
familiar setting of second quantization. This is done in
Sec.~\ref{sec:howe}.

The rest of my article is devoted to the special case of
orthogonal-orthogonal duality. This case is also a main topic of
Ref.~\onlinecite{ref:Nee19}, where, in particular, I prove an
orthogonal-orthogonal duality theorem pertaining to Lie algebras by a
calculation of characters similar to one used by Helmers more than
half a century ago to obtain an analogous symplectic-symplectic
duality theorem.\cite{ref:Hel61} My present investigation springs from
the observation that the orthogonal-orthogonal case of Howe's theorem
relates equivalence classes of irreducible representations of a group
and a Lie algebra, while mine relates equivalence classes of
irreducible representations of two Lie algebras. This makes a
difference because the representations of an orthogonal group and its
Lie algebra have different reducibilities. The distinguishing feature
of the orthogonal groups which gives rise to this difference is the
presence of reflections, which form a coset topologically disconnected
from the subgroup of rotations. An analysis of the representation of
reflections therefore becomes a main theme. The central discussion
appears in Sec.~\ref{sec:o-o&O-o} and leads to the conclusion that one
can prove the orthogonal-orthogonal special case of Howe's duality
theorem and explicitly describe the relation between the dual
equivalence classes by combining my theorem in
Ref.~\onlinecite{ref:Nee19} with that analysis.

To prepare this discussion I review in Sec.~\ref{sec:O&o} essential
parts of the representation theory of orthogonal groups and their Lie
algebras and introduce generalized Young diagrams to describe the
equivalence classes of irreducible representations of orthogonal Lie
algebras. The final Sec.~\ref{sec:refl} continues the theme of
reflections. I thus define there a reflection of the number
non-conserving partner in the dual relationship of Lie algebras, which
contributes to a picture of an almost perfect symmetry between this
and its number conserving mate.

For precision of the terminology, the word \textit{irrep} will denote
an \textit{equivalence class} of irreducible representations of a
group or Lie algebra or a standard realization within such a class.
Throughout, the base field is understood to be the field
$\mathbb C$ of complex numbers.

\section{\label{sec:howe}Howe's duality theorem}

Consider a system of different kinds of particles. Some of them may be
kinds of boson and some of them kinds of fermions. They share a
1-particle state space $V$ with a basis $(| p \rangle, p=1,\dots,d)$.
A particle of kind $\tau$ is created from the vacuum in the state
$|p\rangle$ by the operator $a^\dagger_{p\tau}$. These operators and a
set of corresponding annihilation operators $a_{p\tau}$ obey the the
usual commutation relations
\begin{equation}\label{eq:com}
  |a^\dagger_{p\tau},a^\dagger_{q\upsilon}|
  = |a_{p\tau},a_{q\upsilon}| = 0 , \quad
  |a_{p\tau},a^\dagger_{q\upsilon}| = \delta_{p\tau,q\upsilon} ,
\end{equation}
where $|\cdot,\cdot|$ denotes the anticommutator $\{\cdot,\cdot\}$
when both $\tau$ and $\upsilon$ are kinds of fermions and otherwise
the commutator $[\cdot,\cdot]$. Any change of the basis
$(| p \rangle, p=1 , \dots , d)$ is required to preserve the
commutation relations~\eqref{eq:com}. Despite the notation,
$a^\dagger_{p\tau}$ and $a_{p\tau}$ are not assumed Hermitian
conjugates. No Hermitian inner product is defined, indeed, on the
state space. I call the span $A$ of the set of operators
$a^\dagger_{p\tau}$ and $a_{p\tau}$ the space of field operators.

A classical group $G$ is supposed to act on $V$. The classical groups
are: the general linear group $\gGL(d)$, the orthogonal group
$\gO(d)$, and the symplectic group $\gSp(d)$. Each of them is a group
of linear transformations $g$ of $V$. The matrix elements of $g \in G$
in the basis $(| p \rangle, p=1,\dots,d)$ are denoted by
$\langle p | g | q \rangle$. There is an induced representation of $G$
on the associative algebra $\lA$ generated by the field operators.
This representation is such that $g \in G$ acts distributively on any
product of elements $x \in \lA$, so the action of $g$ on $x$ may be
written conveniently as a formal similarity map
$x \mapsto g x g^{-1}$. In particular $1 \in \lA$ is $G$ invariant.
The action of $g$ on field operators preserves the commutation
relations~\eqref{eq:com}. Dependent on the kind $\tau$ of particle it
may be either cogredient,
\begin{equation}
  g a^\dagger_{p\tau} g^{-1}
  = \sum_q a^\dagger_{q\tau} \langle q | g | p \rangle , \quad
  g a_{p\tau} g^{-1}
  = \sum_q \langle p | g^{-1} | q \rangle  a_{q\tau} ,
\end{equation}
or contragredient,
\begin{equation}
  g a^\dagger_{p\tau} g^{-1}
  = \sum_q \langle p | g^{-1} | q \rangle  a^\dagger_{q\tau} , \quad
  g a_{p\tau} g^{-1}
  = \sum_q a_{q\tau} \langle q | g | p \rangle .
\end{equation}
The set of particle kinds with cogredient $G$ action is denoted by $K$
and the set with contragredient $G$ action by $K^*$. Both sets are
finite.

The general linear group $\gGL(d)$ consists of all invertible linear
transformations of $V$, while the subgroups $\gO(d)$ and $\gSp(d)$ are
defined by the conservation of a non-singular, bilinear form $b$ on
$V$,
\begin{equation}
  \sum_{rs} \langle b | r s \rangle 
    \langle r | g | p \rangle  \langle s | g | q \rangle
  = \langle b | p q \rangle \quad \forall g \in G .
\end{equation}
The bilinear form $b$ is symmetric and skew symmetric, respectively,
in the cases of $\gO(d)$ and $\gSp(d)$. It follows that in the case of
$\gSp(d)$, the dimension $d$ is even. A dual bilinear form $b^*$ is
defined by
\begin{equation}
  \sum_{r} \langle b | p r \rangle \langle q r | b^* \rangle 
  = \delta_{pq} .
\end{equation}
It has the same symmetry as $b$ and satisfies
\begin{equation}
  \sum_{rs}  
    \langle p | g | r \rangle  \langle q | g | s \rangle
    \langle r s | b^* \rangle
  = \langle p q | b^* \rangle \quad \forall g \in G .
\end{equation}
A dual basis $(| p^* \rangle, p=1,\dots,d)$ for $V$ may be defined by
\begin{equation}\label{eq:du_ba}
  \langle b | p q^* \rangle = \delta_{pq} .
\end{equation}
If $G$ acts cogrediently in the basis $(| p \rangle, p=1,\dots,d)$ it
acts contragrediently in the basis $(| p^* \rangle, p=1,\dots,d)$ and
vice versa. For $G = \gO(d)$ or $\gSp(d)$ there is therefore no need
of a distinction between co- and contragredient action, so one can set
$K^* = \emptyset$.

The product $|\cdot,\cdot|$ can be extended to the set
\begin{equation}
  \bar \lh = \text{span} \, \{ \,a b \; | \;
     a,b \in A \, \} ,
\end{equation}
which, by the commutation relations~\eqref{eq:com}, includes the
numbers. I thus set $|a b , c d| = [a b , c d]$ when either both $a$
and $b$ or both $c$ and $d$ are boson field operators or both of them
are fermion field operators, and $|a b , c d| = \{a b , c d\}$ when
both $a b$ and $c d$ are products of one boson field operator and one
fermion field operator. One can check that this defines
$|\cdot,\cdot|$ unambiguously as a bilinear product on $\bar \lh$ and
that $\bar \lh$ is closed under the action of $|\cdot,\cdot|$. In
particular $|h_1 , h_2| = [h_1 , h_2] = 0$ when any one of
$h_1,h_2 \in \bar \lh$ is a number. The algebra
$(\bar \lh , |\cdot,\cdot|)$ is almost a Lie algebra. In
Ref.~\onlinecite{ref:How89}, Howe calls the subalgebra
$(\lh , |\cdot,\cdot|)$ to be defined in a moment a graded Lie
algebra, referring to a grading modulo 2 where numbers and products of
two boson field operators or two fermion field operators have grade 0
and products of a boson field operator and a fermion field operator
have grade 1. In the terminology of Jacobson,\cite{ref:Jac62}
$(\bar \lh , |\cdot,\cdot|)$ is a weakly closed subset of the
associative algebra $\lA$. The set $\bar \lh$ has a subset
\begin{equation}
  \lh = \text{span} \{ \, ]a , b[ \; | \;
    a,b \in A \, \} ,
\end{equation}
where the bilinear product $]\cdot,\cdot[$ is the exact opposite of
the product $|\cdot,\cdot|$ in the sense that $]a , b[ = \{a , b\}$
when either $a$ or $b$ is a boson field operator, and
$]a , b[ = [a , b]$ when both of them are fermion field operators. One
can check that $(\lh , |\cdot,\cdot|)$ is a subalgebra of
$(\bar \lh , |\cdot,\cdot|)$. One can define also its pointwise $G$
invariant subset
\begin{equation}\label{eq:hG}
  \lh^G = \{ \, h \in \lh \; | \;
    g h g^{-1} = h \quad \forall g \in G \, \} .
\end{equation}
The algebra $(\lh^G , |\cdot,\cdot|)$ is a subalgebra of
$(\lh , |\cdot,\cdot|)$ because the map $x \mapsto g x g^{-1}$
preserves relations expressed by the product $|\cdot,\cdot|$.

Finally, the Fock space $\Phi$ is defined by
\begin{equation}
  \Phi = \lA |\rangle = \lA^\dagger |\rangle
\end{equation}
in terms of the vacuum state $|\rangle$. Here $\lA^\dagger$ denotes
the subalgebra of $\lA$ generated by 1 and the creation operators. An
action of $G$ on $\Phi$ is defined by taking literally the formal
similarity map $x \mapsto g x g^{-1}$ and assuming that the vacuum is
$G$ invariant,
\begin{equation}
  g |\rangle = |\rangle \quad \forall g \in G .
\end{equation}
Clearly, $\Phi$ and $\lA^\dagger$ are isomorphic as vector spaces, so
the action of $\lA$ on $\Phi$ can be seen equivalently as an action on
$\lA^\dagger$. So one avoids introducing the space $\Phi$. This point
of view is taken in Ref.~\onlinecite{ref:How89}.

Given these preliminaries, Howe's duality theorem (Theorem 8 of
Ref.~\onlinecite{ref:How89}) can be formulated as follows.

\begin{thm}[Howe]\label{th:howe}
The Fock space $\Phi$ has a decomposition
\begin{equation}\label{eq:decompH}
   \Phi = \bigoplus_\lambda \textup{X}_\lambda \otimes \Psi_\lambda ,
\end{equation}
where the group $G$ and the algebra $(\lh^G , |\cdot,\cdot|)$ act
on $\Phi$ in such a way that each space $\textup{X}_\lambda$
carries an irreducible representation of $G$, and each space
$\Psi_\lambda$ carries an irreducible representation of
$(\lh^G , |\cdot,\cdot|)$. For $\lambda \ne \mu$, the representations
of $G$ on $\textup{X}_\lambda$ and $\textup{X}_\mu$ are inequivalent,
and the representations of $(\lh^G , |\cdot,\cdot|)$ on $\Psi_\lambda$
and $\Psi_\mu$ are inequivalent. The spaces $\textup{X}_\lambda$ have
finite dimensions.
\end{thm}

\begin{rem}
In the formulation of the theorem in Ref.~\onlinecite{ref:How89},
$\lh^G$ is defined as the set of elements in $(\lh , |\cdot,\cdot|)$
which commute with every elements of an image of $G$'s Lie algebra.
There are cases where this set is larger than the set~\eqref{eq:hG};
see my discussion in Ref.~\onlinecite{ref:Nee19} of the case of
fermions and $G = \gO(2)$. The formulation above corresponds to the
proof in Ref.~\onlinecite{ref:How89}, which is based on the so-called
double commutant theorem and invariant theory.\cite{ref:Wey39}
\end{rem}

Howe's theorem is seen to establish a 1--1 relation between the irreps
carried by $\textup{X}_\lambda$ and $\Psi_\lambda$, but it does not
specify this relation. More specific results in this respect appear in
a later extensive treatise by Howe\cite{ref:How95} and also elsewhere
in the literature. This is discussed in Sec.~\ref{sec:exampl}. The
structure of $\lh^G$ is determined in Ref.~\onlinecite{ref:How89} for
each of the three classical groups $G$. The result can be summarized
as follows.

\medskip

$\lh^{\gGL(d)}$ is spanned by the operators
\begin{equation}\label{eq:hG-GL}
  \sum_p \, ]a^\dagger_{p\tau},a_{p\upsilon}[ ,  \quad
    (\tau,\upsilon) \in K \times K \cup K^* \times K^* , \qquad
  \sum_p \, ]a^\dagger_{p\tau},a^\dagger_{p\upsilon}[ , \quad
  \sum_p \, ]a_{p\tau},a_{p\upsilon}[ , \quad
    (\tau,\upsilon) \in K \times K^* .
\end{equation}

\medskip

$\lh^{\gO(d)}$ and $\lh^{\gSp(d)}$ are spanned by the operators
\begin{equation}\label{eq:hG-O-Sp}
  \sum_p \, ]a^\dagger_{p\tau},a_{p\upsilon}[ ,  \quad
  \sum_{pq} \, ]a^\dagger_{p\tau},a^\dagger_{q\upsilon}[ \,
    \langle p q | b^* \rangle , \quad
  \sum_{pq} \, \langle b | p q \rangle \,
     ]a_{q\tau},a_{p\upsilon}[ , \quad
    (\tau,\upsilon) \in K \times K .
\end{equation}

\medskip

\noindent The operators~\eqref{eq:hG-GL} and \eqref{eq:hG-O-Sp} with
products of two creation or two annihilation operators are not
linearly independent. By the symmetry of $]\cdot,\cdot[$, possibly
combined with that of $b$, they are either symmetric or skew symmetric
in $\tau$ and $\upsilon$. In particular $\tau = \upsilon$ is
prohibited in the case of skew symmetry.

\section{\label{sec:exampl}Examples in physics}

Several special cases of Theorem~\ref{th:howe} were known in physics
before Ref.~\onlinecite{ref:How89}. They concern systems which are
either purely bosonic or purely fermionic so that $\lh^G$ is a Lie
algebra. Also, they do not involve contragredient actions of
$\gGL(d)$, so from now on, I set $K^* = \emptyset$ and define
$k = |K|$.

\subsection{\label{sec:GL-GL}$\gGL(d)$--$\gGL(k)$ duality}

The relations which result in the case $G = \gGL(d)$ were noticed very
early in the history of quantum mechanics. When $K^* = \emptyset$, the
Lie algebra $\lh^{\gGL(d)}$ is spanned by the operators
\begin{equation}\label{eq:gl1}
  \sum_p \, ]a^\dagger_{p\tau},a_{p\upsilon} [
  = 2 \sum_p a^\dagger_{p\tau} a_{p\upsilon}
    \pm  \delta_{\tau\upsilon}  d , \quad
  (\tau,\upsilon) \in K \times K ,
\end{equation}
for boson and fermions. The numeric term in Eq.~\eqref{eq:gl1} makes
no difference with respect to the decomposition of $\Phi$, so
$\lh^{\gGL(d)}$ may be replaced equivalently by the Lie algebra
spanned by the operators
\begin{equation}\label{eq:gl2}
  \sum_p a^\dagger_{p\tau} a_{p\upsilon} , \quad
  (\tau,\upsilon) \in K \times K . 
\end{equation}
This is the Lie algebra of the representation of $\gGL(k)$ on $\Phi$
induced by the group of invertible linear transformations acting on
the index $\tau$ of $a^\dagger_{p\tau}$. Irreps of $\gGL(k)$ stay
irreducible upon restriction to the special linear group $\gSL(k)$ of
transformations with determinant 1 because the transformations in
$\gGL(k)$ deviate from those in $\gSL(k)$ only by numeric factors.
Similarly irreps of $\gGL(k)$'s Lie algebra $\lgl(k)$ stay irreducible
upon restriction to $\gSL(k)$'s Lie algebra $\lsl(k)$ because the
transformations deviate only by numeric terms. Finally $\lsl(k)$
irreps exponentiate to $\gSL(k)$ irreps because $\gSL(k)$ is simply
connected. By combination of these facts it follows that, in the
statement of the dual relation, $\lh^{\gGL(d)}$ may be replacedy
equivalently by the $\gGL(k)$ group of transformations acting on
$\tau$, so the relation may be characterized as a $\gGL(d)$--$\gGL(k)$
duality.

Because the operators~\eqref{eq:gl2} conserve the number $n$ of
particles, the sum~\eqref{eq:decompH} splits into parts $\Phi_n$ with
definite $n$. A state in $\Phi_n$ is described by a wave function
\begin{equation}
  \phi(p_1,\dots,p_n,\tau_1,\dots,\tau_n)
\end{equation}
which satisfies
\begin{equation}
  \phi = \mathcal S \phi
\end{equation}
with
\begin{equation}
  \mathcal S = \frac 1 {n!} \sum_{s \in \gS(n)} \, s_p s_\tau
  \qquad \text{and} \qquad \mathcal S = \frac 1 {n!}
    \sum_{s \in \gS(n)} (\text{sgn} \, s) \, s_p s_\tau
\end{equation}
for bosons and fermions, respectively. Here $\gS(n)$ denotes the group
of permutations of $n$ elements, and $s_p$ and $s_\tau$ the
permutation $s$ applied to the arguments $p_i$ and $\tau_i$ of $\phi$,
respectively. The classical example is the system of $n$ electrons in
an atomic shell with principal and azimuthal quantum numbers $N$ and
$l$. Its wave functions can be written
\begin{equation}\label{eq:phi(m_l,m_s)}
  \phi(m_{l1},\dots,m_{ln},m_{s1},\dots,m_{sn})
\end{equation}
in terms of the spatial and spin magnetic quantum numbers $m_l$ and
$m_s$ of one electron. The groups $\gGL(d)$ and $\gGL(k)$ act on the
spatial and spin variables, respectively, of a single electron, so
$d = 2l + 1$ and $k = 2$. The wave function~\eqref{eq:phi(m_l,m_s)}
can be expanded on products
\begin{equation}\label{eq:chi-psi}
  \chi^{\lambda_l}_{\mu_l \nu_l} (m_{l1},\dots,m_{ln}) \,
  \psi^{\lambda_s}_{\mu_s \nu_s} (m_{s1},\dots,m_{sn}) ,
\end{equation}
where $\lambda_l$ and $\lambda_s$ are $n$-cell Young
diagrams.\cite{ref:You00,ref:You02} For a given $\lambda_l$, the
functions $\chi^{\lambda_l}_{\mu_l \nu_l}$ carry the corresponding
irrep of $\gGL(d) \otimes \gS(n)$ with $\gGL(d)$ and $\gS(n)$ acting
on the indices $\mu_l$ and $\nu_l$ respectively, and similarly
$\psi^{\lambda_s}_{\mu_s \nu_s} (m_{s1},\dots,m_{sn})$ in terms of
$\gGL(k)$. (Here, use is made of a well-known relation between
symmetry and $\gGL(d)$ irrep,\cite{ref:Sch01,ref:Wey39} sometimes
called the Schur or Schur-Weyl duality. That group theory has quantum
mechanical applications was seen and communicated by Weyl almost
immediately following the birth of quantum mechanics in
1925.)\cite{ref:Wey28}

The matrices of $\gS(n)$ irreps can be chosen real
orthogonal.~\cite{ref:Fro03} For a given Young diagram $\lambda$, let
$\tilde \lambda$ denote the conjugate diagram, obtained by reflection
in the bisector of the upper left corner. The $\lambda$ and
$\tilde \lambda$ irreps have equal dimensions, and their carrier
spaces have bases $( | \nu \rangle )$ and $( | \tilde \nu \rangle )$
such that, in an obvious notation,
\begin{equation}
  \langle \nu | s | \nu' \rangle_\lambda
  = \text{sgn} \, s \,
  \langle \tilde \nu | s | \tilde \nu' \rangle_{\tilde \lambda}
  \qquad \forall s \in \gS(n) .
\end{equation}
It then follows by the orthogonality relations of unitary matrix
elements of irreps of finite groups\cite{ref:Fro96,ref:Sch05}
(sometimes called the Schur orthogonality relations) that when the
antisymmetrizer $\mathcal S$ is applied to $\phi$, only the terms with
$\lambda_l = \tilde \lambda_s$ survive in the expansion on
products~\eqref{eq:chi-psi} and that these terms combine to sums
\begin{equation}\label{eq:sum_nu}
  \sum_\nu \chi^{\lambda}_{\mu_l \nu} (m_{l1},\dots,m_{ln}) \,
    \psi^{\tilde \lambda}_{\mu_s {\tilde\nu}} (m_{s1},\dots,m_{sn}) .
\end{equation}
We have thus arrived in this special case at another proof of the
duality theorem, which can obviously be generalized to any pair of
dimensions $d$ and $k$. We have even got a precise relation between
the connected irreps of $\gGL(d)$ and $\gGL(k)$: Their Young diagrams
are mutually conjugate. Moreover, because $\lambda$ and
$\tilde \lambda$ can have no more than $d$ and $k$ rows, respectively
(see Sec.\ref{sec:sl-hw-ten}), they can have no more than $k$ and $d$
columns, respectively. Conversely, the sum~\eqref{eq:sum_nu} can be
constructed for every such pair of conjugate Young diagrams $\lambda$
and $\tilde \lambda$, so all of them occur.

Evidently, an even simpler argument gives an analogous result in the
bosonic case. There, the $\gGL(d)$ and $\gGL(k)$ irreps have the same
Young diagram whose depth does not exceed $\min(d,k)$. In physics, one
usually considers the subgroups $\gU(d)$ and $\gU(k)$ of unitary
transformations rather than the full general linear groups $\gGL(d)$
and $\gGL(k)$, and the $\gGL(d)$--$\gGL(k)$ duality is called a
$\gU(d)$--$\gU(k)$ duality. As the restriction to $\gU(d)$ does not
break the irreducibility of $\gGL(d)$ irreps,\cite{ref:Wey39} this
makes no difference with respect to the decomposition of $\Phi_n$.

\subsection{\label{sec:Sp-Sp}$\gSp(d)$-$\gSp(2 k)$ duality}

It can be checked that $\lh^{\gO(d)}$ has the structure of $\lsp(2 k)$
for boson systems and $\lo(2 k)$ for fermion systems and vice versa
for $\lh^{\gSp(d)}$, where $\lo(2 k)$ and $\lsp(2 k)$ are the Lie
algebras of $\gO(2 k)$ and
$\gSp(2 k)$.\cite{ref:How89,ref:Hel61,ref:Row11,ref:Nee19} For fermion
systems, $\Phi$ has finite dimension. So do then the irreps of
$\lsp(2 k)$ or $\lo(2 k)$ in Eq.~\eqref{eq:decompH}. In particular,
because the group $\gSp(2 k)$ is simply connected, the $\lsp(2 k)$
irreps exponentiate to $\gSp(2 k)$ irreps and the
$\gSp(d)$--$\lsp(2 k)$ duality is equivalent to an
$\gSp(d)$--$\gSp(2 k)$ duality. This special case of
Theorem~\ref{th:howe} was noticed and proved by Helmers in
1961.\cite{ref:Hel61} His background was a line of research initiated
in 1943, when Racah introduced the concept of seniority in an analysis
of the $n$-electron system of Sec.~\ref{sec:GL-GL}.\cite{ref:Rac43}
Racah observed that the states in $\Phi$ can be arranged in sequences
of states of $v$, $v + 2$, $v + 4$, \dots\ electrons such that the
successor of a state in the sequence results when a pair of electrons
with total spatial angular momentum $L = 0$ is added to its
predecessor. The number $v$ of electrons in the first state of the
sequence he called the seniority of the sequence.

In 1949, Racah interpreted this result in terms of group theory,
noticing that the Clebsch-Gordan coefficient
$\langle l m_l l m_l' | 0 0 \rangle$ is the matrix element of a
symmetric bilinear form on the space $V$ of spatial 1-electron
states.\cite{ref:Rac49} This bilinear form defines an orthogonal group
$\gO(d)$, and the seniority is a function of the irrep of this group.
Racah's analysis is, in fact, closely related to Weyl's construction
of $\gO(d)$ irreps (see Sec.~\ref{sec:O&o-ten}).\cite{ref:Wey39} Racah
points out in the same article that one can alternatively define a
bilinear form in terms of the Clebsch-Gordan coefficient
$\langle j m j m' | 0 0 \rangle$, where $j$ is the half-integral
quantum number of total, spatial plus spin, angular momentum of the
electron, and $m$ its associated magnetic quantum number. This
Clebsch-Gordan coefficient is skew symmetric in $m$ and $m'$, and an
analysis in terms a symplectic group $\gSp(d)$ ensues. Working in the
basis of 1-electron states defined by $j$ and $m$ instead of $m_l$ and
$m_s$ is known as $jj$ coupling as opposed to $LS$ coupling.

This analysis was subsequently adapted to the nuclear shell model.
There, the spin variable may be supplemented by the variable of
isospin, giving rise to more complicated structures. Adopting the $jj$
coupling scheme, Helmers cast this entire analysis into a framework of
second quantization and then proved, by a calculation of characters,
the general $\gSp(d)$--$\gSp(2 k)$ duality theorem covering the cases
$k = 1$ (electrons, only neutrons, only protons) and $k = 2$ (neutrons
and protons) as well as any greater number of kinds of fermions.
Helmers's proof provides a precise and somewhat peculiar rule in terms
of Young diagrams for the association of the connected $\gSp(d)$ and
$\gSp(2 k)$ irreps: The $\gSp(d)$ Young diagram and a reflected and
rotated copy of the $\gSp(2 k)$ Young diagram fill a rectangle of
depth $d/2$ and width $k$ without overlap. (See the analogous
orthogonal-orthogonal diagram~\eqref{eq:neer} below.) This rule is given
independently by Howe in Ref.~\onlinecite{ref:How95}.

For $k = 1$ the group $\gSp(2 k)$ was identified by Kerman
simultaneously with and independently of Helmers.\cite{ref:Ker61} More
precisely, Kerman identified the unitary $\lsu(2)$ subalgebra of the
$\lsl(2)$ Lie algebra isomorphic to the Lie algebra $\lsp(2)$ of
$\gSp(2 k)$. This $\lsu(2)$ Lie algebra is known as the quasispin
algebra. The 1--1 relation between quasispin and $jj$ seniority is at
the core of an extensive analysis of the nuclear $k = 1$ system by
Talmi and his coworkers.\cite{ref:Tal93} Also independently of
Helmers, Flowers and Szpikowski identified in 1964 the $\lsp(4)$ Lie
algebra pertaining to the case $k = 2$ as an $\lo(5)$ Lie
algebra.\cite{ref:Flo64a} It is well known that $\lsp(4)$ and $\lo(5)$
are isomorphic. In these works, equivalent expressions for the
eigenvalues of a certain ``pairing'' interaction are shown to result
whether expressed by $\lsp(d)$ or $\lsp(2 k)$ quantum numbers upon a
suitable association of these quantum numbers. There is no proof that
the associated irreps select the same subspace of $\Phi$, nor that the
representation of $\lsp(d) \oplus \lsp(2 k)$ on this subspace is
irreducible.

\subsection{\label{sec:O-o}$\gO(d)$-$\lo(2 k)$ duality}

Helmers anticipates in Ref.~\onlinecite{ref:Hel61} that results
similar to those obtained there hold when the single-fermion angular
momentum quantum number $l$ is integral so that the Clebsch-Gordan
coefficient $\langle l m_l l m_l' | 0 0 \rangle$ provides a symmetric
bilinear form, but the road to such analogous results would turn out
fairly long. Closely following their proposal of the $\lo(5)$ Lie
algebra, Flowers and Szpikowski proposed an $\lo(8)$ Lie algebra of
``quasi-spin in $LS$ coupling'' which is recognized as $\lh^{\gO(d)}$
in the case when $V$ is the space of spatial angular momentum states
as in Sec.~\ref{sec:GL-GL} and $k$ equals 4 corresponding to the
4-dimensional space of the nucleonic spin and
isospin.\cite{ref:Flo64b} Like in the case of $\lo(5)$ they show that
equivalent expressions for the eigenvalues of a pairing interaction
result whether expressed by $\lo(d)$ or $\lo(8)$ quantum numbers upon
a suitable association of these quantum numbers, but there is no proof
of duality in the sense of Theorem~\ref{th:howe}.

Later work in nuclear physics has been based on this suggestion, but
not before the work of Rowe, Repka, and Carvalho to appear in 2011
was, to my knowledge, the relation between the dual irreps of $\gO(d)$
and $\lo(2 k)$ characterized precisely.\cite{ref:Row11} (The case
$k = 4$ was handled by Rowe and Carvalho in 2007.)\cite{ref:Row07} In
particular, unlike the case of the \mbox{$\gSp(d)$--$\gSp(2k)$}
duality, Howe does not in Ref.~\onlinecite{ref:How95} provide such a
precise characterization in the $\gO(d)$--$\lo(2k)$ case. The argument
in Ref.~\onlinecite{ref:Row11} is based on the identification of a
state in $\Phi$ that has highest weight simultaneously with respect to
$\lo(d)$, $\lsl(d)$, $\lsl(k)$, and $\lo(2 k)$. (See
Eq.~\eqref{eq:phi_hw}.) The $\gGL(d)$--$\gGL(k)$ duality relation then
provides a relation between the $\gO(d)$ and $\lo(2 k)$ irreps. In
2019 I used Helmers's method to obtain a result that is closely
related to but slightly different from those of Howe in
Ref.~\onlinecite{ref:How89} and Rowe, Repka, and Carvalho in
Ref.~\onlinecite{ref:Row11} by establishing a relation between irreps
of the Lie algebras $\lo(d)$ and $\lo(2 k)$ rather than between such
of the group $\gO(d)$ and the Lie algebra $\lo(2 k)$.\cite{ref:Nee19}
This makes a difference because, consistently with the
non-connectedness of $\gO(d)$, representations of $\gO(d)$ and
$\lo(d)$ have different reducibilities.~\cite{ref:Wey39} The relation
between my result and those of Refs.~\onlinecite{ref:How89,ref:Row11}
is the topic of Sec.~\ref{sec:o-o&O-o}.

\subsection{\label{sec:O-sp&Sp-o}$\gO(d)$-$\lsp(2 k)$ and
  $\gSp(d)$-$\lo(2 k)$ dualities}

In the cases of the $\gO(d)$-$\lsp(2 k)$ and $\gSp(d)$-$\lo(2 k)$
dualities, the highest weight of an $\lh^G$ irrep in the
decomposition~\eqref{eq:decompH} can be expressed by that of the $G$
irrep by the method of Rowe, Repka, and Carvalho.~\cite{ref:Row11} A
Borel subalgebra $\lb$ of $\lh^G$ is spanned by the subset
\begin{equation}
  2 \sum_p a^\dagger_{p\tau} a_{p\upsilon} + d ,  \quad
    \tau \le \upsilon , \qquad
  2 \sum_{pq} \, \langle b | p q \rangle a_{p\tau} a_{q\upsilon} ,
\end{equation}
of the set~\eqref{eq:hG-O-Sp}. The members of its Cartan subalgebra
whose eigenvalues on the highest-weight vector defined by $\lb$ give
in any finite-dimensional irreducible representation the row lengths
of its Young diagram (compare~Sec.~\ref{sec:spin}) are
\begin{equation}\label{eq:Cartan}
  - \sum_p  a^\dagger_{p\tau} a_{p\tau} - d/2 , \quad
    \tau = 1 , \dots , k .
\end{equation}
By the discussion in Sec.~\ref{sec:GL-GL}, the number of rows in the
$G$ irrep's Young diagram cannot exceed $\min(d,k)$. Let
$\lambda_p, p = 1 , \dots , k$, denote their lengths in the order from
top to bottom with trailing zeros if their number is less than $k$.
The highest weight of the $\lh^G$ irrep is then given in terms of the
eigenvalues $w_\tau$ of the operators~\eqref{eq:Cartan} by
\begin{equation}\label{eq:w_tau}
  w_\tau = - \lambda_{k+1-\tau} - d/2 .
\end{equation}
With every $w_\tau$ negative, this cannot be a linear combination of
fundamental weights with non-negative integral coefficients unless
$\lh^G \simeq \lo(2)$, so in any other case the $\lh^G$ irrep has
infinite dimension.\cite{ref:Jac62}

(The Lie algebra $\lo(2)$ is 1-dimensional and thus has a continuum of
1-dimensional irreps. Since $k = 1$, we are dealing with a single kind
of bosons. For the skew symmetric bilinear form $b$ pertaining to
$\gSp(d)$, the operators~\eqref{eq:hG-O-Sp} with products of two
creation or two annihilation operators are then absent, whence
$\lo(2)$ conserves the number $n$ of bosons. In fact $\lo(2)$ is
spanned in its present realization by the single remaining
operator~\eqref{eq:hG-O-Sp}. The representation of $\gGL(d)$ on
$\Phi_n$ is easily seen to belong to the irrep with a 1-row Young
diagram of width $n$. The states in $\Phi_n$ are also seen to be
traceless with respect to $b$ (compare Sec.~\ref{sec:O&o-ten}), so the
representation stays irreducible upon restriction to $\gSp(d)$, whence
$\lambda_1 = n$. The eigenvalue~\eqref{eq:w_tau} of the single
operator~\eqref{eq:Cartan} is then tautological. The
decomposition~\eqref{eq:decompH} is therefore, in this pathological
case, just the splitting of $\Phi$ into its subspaces $\Phi_n$.)

Rowe, Repka, and Carvalho derive in Ref.~\onlinecite{ref:Row11} the
relation~\eqref{eq:w_tau} for $G = \gO(d)$. While their proof rests
entirely on properties of the Lie algebra $\lsp(2 k)$, they describe
their result as an $\gO(d)$--$\gSp(k,\mathbb R)$ duality, where
$\gSp(k)$ in their notation is $\gSp(2 k)$ in mine. At the level of
Lie algebras the restriction to the reals is irrelevant. One can
chose, however, in $A$ a basis of coordinates and momenta
\begin{equation}
  X_{p\tau} = c (a^\dagger_{p\tau} + a_{p\tau}) , \quad
  P_{p\tau} = \frac i {2 c} (a^\dagger_{p\tau} - a_{p\tau}) ,
\end{equation}
with $c$ a numeric constant. These operators obey the Heisenberg
commutation relations
\begin{equation}
  [ X_{p\tau} , P_{q\upsilon} ] = i \, \delta_{p\tau,q\upsilon} ,
\end{equation}
and $\lsp(2 k)$ is spanned by the operators
\begin{equation}\label{eq:can}
  i \sum_p X_{p\tau} X_{p\upsilon} , \quad
  i \sum_p P_{p\tau} P_{p\upsilon} , \quad
  i \sum_p \{ X_{p\tau} , P_{p\upsilon} \} .
\end{equation}
A real linear combination $M$ of the operators~\eqref{eq:can} induces
by the commutation map $x \mapsto [M , x]$ an infinitesimal linear
canonical transformation among the coordinates $X_{p\tau}$ and momenta
$P_{p\tau}$ pertaining to a fixed $p$, and in classical mechanics the
linear canonical transformations in $k$ dimensions form the group
$\gSp(2 k,\mathbb R)$. This has a double covering group, the so-called
metaplectic group $\gMp(2 k)$, faithfully represented on a boson Fock
space,\cite{ref:Sha62} and the authors of Ref.~\onlinecite{ref:Row11}
suggest that the $\lsp(2 k,\mathbb R)$ irrep with the highest
weight~\eqref{eq:w_tau} may in general exponentiate to an irrep of
$\gMp(2 k)$. An overlapping group of authors proposed a model of
nuclear collective motion based on the Lie algebra $\lsp(6)$,
corresponding to $k = 3$, where $p$ essentially labels the nucleons
and $\tau$ the three spatial dimensions.\cite{ref:Ros77} The bosons
are quanta of oscillation in the conventional harmonic oscillator
potential well of the nuclear shell model. In the literature on this
model, $\lsp(6)$ is called $\gSp(3,\mathbb R)$.

In a study in the framework of the so-called interacting boson
model of nuclear spectra,\cite{ref:Iac80} where correlated pairs of
nucleons are represented by bosons, Lerma~\textit{et~al.} employ the
real form $\lo(3,2)$ (called $\gSO(3,2)$) of the $\lsp(4)$ Lie algebra
dual to $\gO(d)$ for $k = 2$.\cite{ref:Ler06} Like in the model of
Flowers and Szpikowski discussed in Sec~\ref{sec:Sp-Sp}, the two
values of $\tau$ correspond to neutrons and protons. The space $V$ is
spanned by boson states common to both kinds of nucleon.

\section{\label{sec:O&o}Finite-dimensional representations of $\gO(d)$
  and $\lo(d)$}

In this section, I sketch some elements of the representation theory
of the orthogonal groups and their Lie
algebras.\cite{ref:Wey39,ref:Jac62,ref:Goo98} At the end, I introduce
convenient Young diagrams for the description of finite-dimensional
$\lo(d)$ irreps.

\subsection{\label{sec:sl-hw-ten}$\lsl(d)$ highest-weight vectors on a
  tensor space}

Already met are the functions
\begin{equation}\label{eq:chi_la_mu_nu}
  \chi^\lambda_{\mu \nu} (p_1 , \dots , p_n) ,
\end{equation}
which carry irreducible representations of $\gGL(d) \otimes \gS(n)$
and span the space $T_n(d)$ of functions (or tensors)
\begin{equation}
  \chi (p_1 , \dots , p_n) .
\end{equation}
The action of $g \in \gGL(d)$ on $\chi$ is by
\begin{equation}\label{eq:q-action}
  (g \chi) (p_1 , \dots , p_n)
  = \sum_{(q_1,\dots,q_n)}
      \left(\prod_i \langle p_i | g | q_i \rangle \right)
      \chi (q_1 , \dots , q_n) ,
\end{equation}
and the action of $s \in \gS(n)$ by permutation of the arguments
$p_i$. Here, $n \ge 1$. One can include $n = 0$ by defining an empty
tuple $()$ invariant to a group $\gS(0) = \{ 1 \}$ and assigning the
value 1 to an empty product. Then $T_0(d)$ is 1-dimensional and
carries the 1 representation of
$\gGL(d) \otimes \gS(0) \simeq \gGL(d)$. The corresponding Young
diagram $\lambda$ is the empty diagram. For fixed $\lambda$ and $\nu$
the functions~\eqref{eq:chi_la_mu_nu} form a basis for an irreducible
$\gGL(d)$ module. Every equivalent module results from this one by the
action of a member of the $\gS(n)$ group algebra. Fixing $\lambda$ and
$\nu$ amounts to projecting $T_n(d)$ by one of the primitive
idempotents associated with $\lambda$ within the $\gS(n)$ group
algebra.\cite{ref:Wey39} This idempotent can be chosen as the Young
symmetrizer
\begin{equation}\label{eq:Y}
  Y_\lambda = c_\lambda \sum_{st} (\text{sgn} \, s) st 
\end{equation}
corresponding to the tableau where the numbers $1, \dots, n$ are
placed in reading order in the cells of the Young diagram $\lambda$.
In the sum in Eq.~\eqref{eq:Y}, the permutation $s$ runs over all
products of permutations of the columns of the tableau, $t$ runs over
all products of permutations of the rows of the tableau, and
$c_\lambda \ne 0$ is a numeric factor ensuring
$Y_\lambda^2 = Y_\lambda$. Consider the function
\begin{equation}\label{eq:chi_Y}
  \chi_\lambda (p_1 , \dots , p_n) = \prod_i \delta_{p_i\rho(i)} ,
\end{equation}
where $\rho(i)$ is the ordinal number of the row in the tableau that
contains the number $i$. This product vanishes if $\lambda$ has more
than $d$ rows. Otherwise, when $Y_\lambda$ acts on $\chi_\lambda$,
every $t$ in the sum~\eqref{eq:Y} fixes $\chi_\lambda$, and every $s$
gives a different product of Kronecker deltas, so
$\chi^\lambda_\text{hw} = Y_\lambda \chi_\lambda \ne 0$. The entire
irreducible $\gGL(d)$ module is then generated from
$\chi^\lambda_\text{hw}$ by the $\gGL(d)$ group algebra.

The space of linear transformations of $V$ is spanned by the
transformations $e_{pq}$ with matrix elements
\begin{equation}
  \langle r | e_{pq} | s \rangle  = \delta_{rs,pq} .
\end{equation}
The action of $x \in \lgl(d)$ on  $\chi \in T_n(d)$ is by
\begin{equation}\label{eq:x-action}
  (x \chi) (p_1 , \dots , p_n)
  = \sum_{(q_1,\dots,q_n)}
      \left(\sum_i \langle p_i | x | q_i \rangle \right)
      \chi (q_1 , \dots , q_n)
\end{equation}
with 0 assigned to an empty sum. For $x = e_{pq}$ and
$\chi = \chi_\lambda$ this gives a sum of terms where one Kronecker
delta $\delta_{p_iq}$ in the product~\eqref{eq:chi_Y} is replaced by
$\delta_{p_i p}$. For $p = q$ this changes nothing and because
$Y_\lambda$ commutes with the action~\eqref{eq:x-action}, one gets
\begin{equation}\label{eq:e=rows}
  e_{pp} \, \chi^\lambda_\text{hw}
  = \lambda_p \, \chi^\lambda_\text{hw},
\end{equation}
where $\lambda_p,p =1,\dots,d$, are the, possibly vanishing, row
lengths of the Young diagram $\lambda$ ordered from top to bottom. For
$p < q$ the factor $\delta_{p_i q}$ becomes identical to a factor
$\delta_{p_j p}$ where, after a permutation $t$ in the
sum~\eqref{eq:Y}, the number $j$ is situated vertically above $i$ in
the Young tableau. The term is then killed by the summation over the
permutations~$s$, so one gets
\begin{equation}\label{eq:sl-hw}
  e_{pq} \, \chi^\lambda_\text{hw} = 0 .
\end{equation}

For $d \ge 2$ the transformations
$e_{pp} - e_{p+1,p+1}, p = 1, \dots , d - 1$, and $e_{pq}, p < q,$
span a Borel subalgebra $\lb$ of $\lsl(d) \subset \lgl(d)$. Its
derived Lie algebra $\lb' = [\lb , \lb]$ is spanned by the second set
of transformations, so $\lb' \chi^\lambda_\text{hw} = 0$ by
Eq.~\eqref{eq:sl-hw}. Thus $\chi^\lambda_\text{hw}$ is a
highest-weight function for the representation of $\lsl(d)$ on
$T_n(d)$. For $d = 1$, we have $\lsl(d) = \lsl(1) = \{ 0 \}$, which is
a particularly trivial case of an Abelian Lie algebra. Other cases met
below include the 1-dimensional Lie algebra $\lo(2)$. The irreps of
Abelian Lie algebras are 1-dimensional. In the present case the
1-dimensional space $T_n(1)$ carries the representation $0 \mapsto 0$
of $\lsl(1)$. An Abelian Lie algebra is a Cartan and a Borel
subalgebra of itself, and any irreducible representation is, being,
since 1-dimensional, isomorphic to a linear form, a weight relative to
this Cartan subalgebra and indeed the only weight of the
representation. So the representation itself may be considered a
highest weight, and any vector in its carrier space a highest-weight
vector. In particular the function $\chi^\lambda_\text{hw}$ with
$\lambda$ the 1-row, $n$-cell Young diagram may be considered a
highest-weight function for the representation of $\lsl(1)$ on
$T_n(1)$.

\subsection{\label{sec:O&o-ten}Irreducible representations of $\gO(d)$
  and $\lo(d)$ on a tensor space}
Let $m $ be some non-negative integer. Weyl shows~\cite{ref:Wey39}
that every irreducible module over $\gO(d) \subset \gGL(d)$ in
$T_m(d)$ is isomorphic for some $n \le m$ to a module in the space
$T^0_n(d)$ of functions $\chi \in T_n(d)$ that are traceless in the
sense
\begin{equation}
  \sum_{p_i p_j} \langle b | p_i p_j \rangle
    \chi (p_1 , \dots , p_n) = 0 \quad \forall i, j, i \ne j .
\end{equation}
The module in $T^0_n(d)$ is embedded in an irreducible $\gGL(d)$
module in $T_n(d)$, and $\gO(d)$ modules embedded in this manner in
inequivalent $\gGL(d)$ modules are inequivalent. The Young diagram of
the $\gGL(d)$ irrep may then be assigned to the $\gO(d)$ irrep. A
$\gGL(d)$ irrep can contain an $\gO(d)$ irrep in this way if and only
if the sum of depths of any two different columns of its Young diagram
does not exceed $d$. The generating function $\chi^\lambda_\text{hw}$
as defined in Sec.~\ref{sec:sl-hw-ten} of such an $\gGL(d)$ module is
seen to be traceless when the bilinear form $b$ is chosen in the form
\begin{equation}\label{eq:b}
  \langle b | p q \rangle = \delta_{p+q,d+1} ,
\end{equation}
so it then also generates the embedded $\gO(d)$ module. Each allowed
Young diagram has a partner, which I call its complementary Young
diagram. Complementary Young diagrams are identical except for the
depths $\tilde \lambda_1$ and $\tilde \lambda_1'$ of their first
columns, which obey $\tilde \lambda_1 + \tilde \lambda_1' = d$. The
matrices of a pair of $\gO(d)$ irreps with different, complementary
Young diagrams can be chosen to coincide for rotations $g \in \gO(d)$,
that is, $\det g = 1$, and differ by a factor $-1$ for reflections,
that is, $\det g = -1$. These representations therefore become
identical upon restriction to the subgroup $\gSO(d)$ of rotations, and
this representation can be shown to be irreducible. (The
irreducibility follows from that of the derived represention of its
Lie algebra $\lo(d)$, which can be inferred from the general theory of
finite-dimensional representations of semisimple Lie algebras
mentioned in Sec.~\ref{sec:spin}.) The $\gSO(d)$ coset of reflections
is generated by its element
\begin{equation}\label{eq:r}
  r = \begin{cases}
    - e_{(d+1)/2,(d+1)/2} + \sum_{p \ne (d+1)/2} e_{pp} ,
    & \text{odd $d$} , \\
    e_{d/2,d/2+1} + e_{d/2+1,d/2} + \sum_{p \ne d/2,d/2+1} e_{pp} ,
    & \text{even $d$} .
  \end{cases}
\end{equation}
It is seen that $r \chi^\lambda_\text{hw} = \chi^\lambda_\text{hw}$
for $\tilde \lambda_1 < d/2$ and
$r \chi^\lambda_\text{hw} = - \chi^\lambda_\text{hw}$ for
$\tilde \lambda_1 > d/2$. An $\gO(d)$ irrep with a self-complementary
Young diagram breaks into two $\gSO(d)$ irreps connected by the
reflections. Self-complementary Young diagrams only occur for even
$d$.

The Lie algebra $\lo(d)$ is that of the maximal connected subgroup
$\gSO(d)$ of $\gO(d)$, so the $\lo(d)$ irreps occurring on
$\bigoplus_n T_n(d)$ are those of $\gSO(d)$. Let transformations
$\bar e_{pq}$ be given by
\begin{equation}\label{eq:ebar}
  \bar e_{pq} = e_{pq} - e_{q^*p^*}
\end{equation}
in terms of the dual basis defined by Eq. \eqref{eq:du_ba}.
Explicitly, $p^* = d + 1 - p$. Also, assume for the moment that
$d \ge 2$. The transformations $\bar e_{pq}, p \le q,$ then span a
Borel subalgebra of $\lo(d)$. Because these transformations belong to
the $\lsl(d)$ Borel subalgebra defined in Sec.~\ref{sec:sl-hw-ten},
the $\lsl(d)$ highest-weight function $\chi^\lambda_\text{hw}$ is also
an $\lo(d)$ highest-weight function. The transformations
$\bar e_{pp}, p = 1, \dots , \lfloor d/2 \rfloor,$ form a basis for
the Cartan subalgebra of this Borel subalgebra. If $\lambda$ and
$\lambda'$ are complementary and
$\tilde \lambda_1 < \tilde \lambda_1'$, these transformations are seen
from Eq.~\eqref{eq:e=rows} to have on $\chi^\lambda_\text{hw}$ and
$\chi^{\lambda'}_\text{hw}$ the same set of eigenvalues by the
action~\eqref{eq:x-action}, equal to the lengths of the first
$\lfloor d/2 \rfloor$ rows of $\lambda$. If $\lambda$ is
self-complementary, let $\lambda'$ be the Young diagram obtained from
$\lambda$ by moving row number $d/2$ one step down (thus violating the
rule that $\lambda_p, p = 1, \dots, d$, should be a non-increasing
sequence). Because every transformation $x$ in $\lo(d)$ satisfies
$\langle d/2 | x | d/2 + 1 \rangle = 0$, the function
$\chi^{\lambda'}_\text{hw}$ is also an $\lo(d)$ highest-weight
function. These $\lo(d)$ highest-weight functions belong to a common
irreducible $\gO(d)$ module because
$\chi^{\lambda'}_\text{hw} = r \chi^{\lambda}_\text{hw}$, and this
$\gO(d)$ module must be the one generated by $\chi^\lambda_\text{hw}$.
Thus $\chi^{\lambda}_\text{hw}$ and $\chi^{\lambda'}_\text{hw}$
generate the two irreducible $\lo(d)$ modules branching out from the
$\gO(d)$ module at the restriction to $\gSO(d)$. The eigenvalues of
$\bar e_{pp}, p = 1, \dots , d/2$, on $\chi^{\lambda'}_\text{hw}$ by
the action~\eqref{eq:x-action} are the same as on
$\chi^{\lambda}_\text{hw}$ except for a change of sign of the
eigenvalue of $\bar e_{d/2,d/2}$. In summary, for $d \ge 2$ the
eigenvalues $w_p$ of
$\bar e_{pp}, p = 1, \dots , \lfloor d/2 \rfloor$, acting
by~\eqref{eq:x-action} on a highest-weight function of an $\lo(d)$
module in $\bigoplus_n T_n(d)$ are integral and satisfy
\begin{equation}\label{eq:w_rules}
  \begin{array}{ll}
    w_1 \ge w_2 \ge \dots \ge w_{(d-1)/2} \ge 0 , & 
      \text{\quad odd $d$} , \\
    w_1 \ge w_2 \ge \dots \ge |w_{d/2}| \ge 0 , &
      \text{\quad even $d$} ,
  \end{array}
\end{equation}
and for every set of integers $w_p$ which obey these rules there is an
embedding $\gO(d)$ module whose Young diagram has row lengths
$\lambda_p = | w_p |$ for $p \le d/2$ and $\lambda_p = 0$ for
$p > d/2$ as well as one with the complementary Young diagram, if
different.

Turning to the case $d = 1$, one has $T^0_n(1) = T_n(1)$ for $n = 0$
or 1 and $T^0_n(1) = 0$ for $n \ge 2$. The representations of
$\lo(1) = \lsl(1) = \{ 0 \}$ on $T^0_0(1)$ and $T^0_1(1)$ are the same
as those of $\lsl(1)$. Consistently with the complementarity of the
0-cell and 1-cell Young diagrams they are equivalent. Both functions
$\chi^\lambda_\text{hw}$ defined by these two Young diagrams are
highest-weight functions. The sequence
$w_p, p = 1, \dots , \lfloor d/2 \rfloor$, is empty, as is the
sequence $\lambda_p, p = 1, \dots , \lfloor d/2 \rfloor$, of row
lengths of the empty Young diagram. In this way the case $d =1$
conforms to the general rule.

\subsection{\label{sec:spin}Spin representations and $\lo(d)$ Young
  diagrams.}

Not every finite-dimensional $\lo(d)$ irrep occurs in
$\bigoplus_n T_n(d)$. There is a general theory which determines every
finite-dimensional irrep of a semisimple Lie
algebra.~\cite{ref:Car13,ref:Jac62} Among the Lie algebras $\lo(d)$
this excludes $\lo(2)$, which is not semisimple. Otherwise, in terms
of the eigenvalues $w_p$ of the transformations
$\bar e_{pp}, p = 1 , \dots , \lfloor d/2 \rfloor $, on the
highest-weight vector determined by the Borel subalgebra spanned by
the transformations $\bar e_{pq}, p \le q$, the result is that an
$\lo(d)$ irrep has finite dimension if and only if either all $w_p$
are integral or all $w_p$ are half-integral and the
rule~\eqref{eq:w_rules} is obeyed. Incidentally, this is also
fulfilled for every $\lo(2)$ irrep met below. It may be considered
conformance to the general rule that the trivial Lie algebra
$\lo(1) = \{ 0 \}$ has only the trivial irrep $0 \mapsto 0$, which may
formally be assigned the empty sequence
$w_p, p = 1 , \dots , \lfloor d/2 \rfloor$ with $d = 1$.

Besides the $\lo(d)$ irreps which occur in $\bigoplus_n T_n(d)$ there
is thus a whole set with half-integral $w_p$, the so-called spin
irreps. It is convenient to define Young diagrams to describe the
entire set of finite-dimensional $\lo(d)$ irreps. In view of the close
relation between the set of $w_p$ and the row lengths of the Young
diagram of an $\gO(d)$ module embedding an $\lo(d)$ module in
$\bigoplus_n T_n(d)$ when such one exists, it is natural to let the
$\lo(d)$ Young diagrams have rows of lengths $w_p$. Unlike the
$\gO(d)$ Young diagrams, the depth of an $\lo(d)$ Young diagram then
does not exceed $d/2$. To describe the negative $w_{d/2}$ which can
occur for even $d$, one must include rows of negative length. For
example the $\lo(6)$ irrep $(w_1,w_2,w_3) = (5,3,-2)$ may be described
by the diagram:
\begin{equation}
  \includegraphics{young14}
\end{equation}
(To distinguish, for $d = 2$, positive and negative $w_1$ from one
another one must mark somehow the edge whence the row extends.) Spin
irreps may be described by the inclusion of a column of width $1/2$,
for example:
\begin{equation}
  \includegraphics{young15}\qquad\includegraphics{young16}
\end{equation}
for the irreps $(w_1,w_2,w_3) = (9/2,5/2,3/2)$ and
$(w_1,w_2,w_3) = (9/2,5/2,-3/2)$, respectively.y

\section{\label{sec:o-o&O-o}$\lo(d)$--$\lo(2 k)$ and
  $\gO(d)$--$\lo(2 k)$ dualities}

After all these preparations, I finally get to my task. In
Ref.~\onlinecite{ref:Nee19}, I prove Theorem~2 below by a
calculation of characters similar to that of Helmers in
Ref.~\onlinecite{ref:Hel61}. It refers to Young diagrams of irreps of
orthogonal Lie algebras as defined in Sec.~\ref{sec:spin} with
$\lambda$ and $w$ describing irreps of $\lo(d)$ and $\lo(2 k)$,
respectively. Their row lengths are denotes by $\lambda_p$ and
$w_\tau$. Irreps of $\lo(d)$ with Young diagrams $\lambda$ that are
identical except for opposite, non-zero, values of $\lambda_{d/2}$
(which only occurs for even $d$), and irreps of $\lo(2 k)$ with Young
diagrams that are identical except for opposite, non-zero, values of
$w_k$, are paired so that the Young diagram with positive length of
this row represents the pair. With this convention, the theorem reads
as follows.

\begin{thm}[Neerg\aa rd]\label{th:neer}
The fermion Fock space $\Phi$ has the decomposition
\begin{equation}\label{eq:decompN}
   \Phi = \bigoplus \textup{X}_\lambda \otimes \Psi_w ,
\end{equation}
where $\lo(d)$ and $\lo(2 k)$ act on $\Phi$ in such a way that 
$\textup{X}_\lambda$ and $\Psi_w$ carry the irreps or pairs of irreps
of $\lo(d)$ and $\lo(2 k)$ with Young diagrams $\lambda$ and $w$, and
the sum runs over all pairs of $\lambda$ and $w$ which fill a
$d/2 \times k$ frame without overlap:
\begin{equation}\label{eq:neer}
  \includegraphics{young17}
\end{equation}
\end{thm}

\noindent The illustration graphically represents, for $d = 13$ and
$k = 4$, a term in the sum~\eqref{eq:decompN} where
$\textup{X}_\lambda$ is an $\lo(d)$ module belonging to the irrep with
Young diagram $\lambda = (4,3,3,2,1,0)$, and $\Psi_w$ an $\lo(2 k)$
module equivalent to the direct sum of the irreps with Young diagrams
$w = (11/2,7/2,5/2,3/2)$, and $w = (11/2,7/2,5/2,-3/2)$. The Young
diagram $w$ is reflected and rotated so that its rows appear
vertically from the right to the left. I deliberately chose an example
with an odd $d$ to illustrate that the $\lo(2 k)$ irreps are spin
representations when $d$ is odd such as in the single-$l$ shell
systems of Secs.~\ref{sec:Sp-Sp} and \ref{sec:O-o}. Because $k$ is
integral, the $\lo(d)$ irrep is always a non-spin representation, as
it should be because $\lo(d)$ is the Lie algebra of an $\gO(d)$ group
acting on $\Phi$. For this reason, each irreducible $\lo(d)$ module in
$\Phi$ is also known to have an embedding $\gO(d)$ module.

Theorem~\ref{th:neer} is seen to be closely parallel to Helmers's
$\gSp(d)$--$\gSp(2 k)$ (or $\lsp(d)$--$\lsp(2 k)$) duality
theorem.~\cite{ref:Hel61} Both are symmetric in the spaces of
dimensions $d$ and $2 k$, and the rules for the association of
diagrams are identical. It may be noticed that always one $\lo(d)$
irrep is associated with a pair of $\lo(2 k)$ irreps and vice
versa. Thus, if the border in Fig.~\eqref{eq:neer} between
$\lambda$ and $w$ hits the bottom edge of the frame, two $\lo(d)$
irreps correspond to one $\lo(2 k)$ irrep. If it hits the left edge,
it is opposite. The first case actually provides a partial proof of
the $\gO(d)$--$\lo(2 k)$ case of Theorem~\ref{th:howe} and also
specifies the relation between the $\gO(d)$ and $\lo(2 k)$ irreps in
this case. Indeed, if the border between $\lambda$ and $w$ hits the
bottom edge of the frame, the $\lo(d)$ Young diagram is
self-complementary, so it represents two different $\lo(d)$ irreps
embedded in an $\gO(d)$ irrep determined by the $\lo(2 k)$ irrep. Let
$\Phi_\psi$ denote the subspace of $\Phi$ selected by a vector
$\psi \in \Psi_w$. By Theorem~\ref{th:neer}, $\Phi_\psi$ is composed
of two irreducible $\lo(d)$ modules belonging to different irreps.
Each $\lo(d)$ module is embedded in an $\gO(d)$ module which also
contains an $\lo(d)$ module belonging to the other irrep, and since
$\gO(d)$ and $\lo(2 k)$ commute acting on $\Phi$, each entire $\gO(d)$
module lies within $\Phi_\psi$. But because each $\lo(d)$ irrep
appears just once in $\Phi_\psi$, the $\gO(d)$ modules must then
coincide.

If the border between $\lambda$ and $w$ hits the left edge of the
frame, as happens in Fig~\eqref{eq:neer}, and as it always does when
$d$ is odd, the correspondence between
the irreps of $\lo(2 k)$ and $\gO(d)$ is less unique. Two $\lo(2 k)$
irreps correspond to the same $\lo(d)$ irrep, and each corresponding
module over $\lo(d)$ in $\Phi$ carries an extension to an $\gO(d)$
module. But the $\gO(d)$ representation may belong to any one of two
irreps, so based on Theorem~\ref{th:neer} alone the range of possible
$\gO(d)$ irreps cannot be narrowed further than to those two.
Theorem~\ref{th:howe} tells us that the $\gO(d)$ irreps corresponding
to the two $\lo(2 k)$ irreps must be different, leaving still two
alternatives for the precise association.

Rowe, Repka, and Carvalho derive the relation
\begin{equation}\label{eq:w-lambda}
  w_\tau = d/2 - \tilde \lambda_{k+1-\tau} ,
\end{equation}
where $\tilde \lambda_\tau$ are the column depths of the $\gO(d)$
Young diagram and $w_\tau$ the row lengths of the $\lo(2 k)$ Young
diagram.\cite{ref:Row11} (Somewhat imprecisely, the Lie algebra
$\lo(2 k)$ is called $\gSO(2 k)$ in Ref.~\onlinecite{ref:Row11}.) This
can be illustrated as follows.
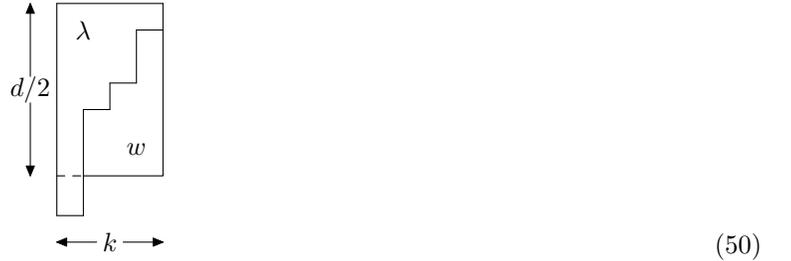
\begin{equation}\label{eq:rowe}
  \includegraphics{young18}
\end{equation}
Here, $d = 13$, $k = 4$, $\lambda = (4,3,3,2,1,1,1,1,0,0,0,0,0)$, and
$w = (11/2,7/2,5/2,-3/2)$, and $w$ is reflected and rotated as in
Fig.~\eqref{eq:neer}. The Young diagrams $\lambda$ and $w$ fill the
$d/2 \times k$ frame without overlap provided a negative $w_k$ is
understood to cancel a part of $\lambda$ which extrudes the frame.
Theorem~5 of Ref.~\onlinecite{ref:Row11} may be formulated as the
statement that every pair of $\gO(d)$ and $\lo(2 k)$ irreps whose
Young diagrams combine in this manner, and only those, occur exactly
once in the decomposition of $\Phi$. The authors of
Ref.~\onlinecite{ref:Row11} get Eq.~\eqref{eq:w-lambda} by associating
a $\gGL(d)$ irrep embedding the $\gO(d)$ irrep with a $\gGL(k)$ irrep
embedded in the $\lo(2k)$ irrep. I shall obtain it more directly by
using Theorem~\ref{th:neer} and analyzing the action on $\Phi$ of the
reflection $r$ given by Eq.~\eqref{eq:r}.

I must prove that when the border in Fig.~\eqref{eq:neer} between
$\lambda$ and $w$ hits the left edge of the frame, as it happens in
the illustration, the pair of $\gO(d)$ Young diagrams
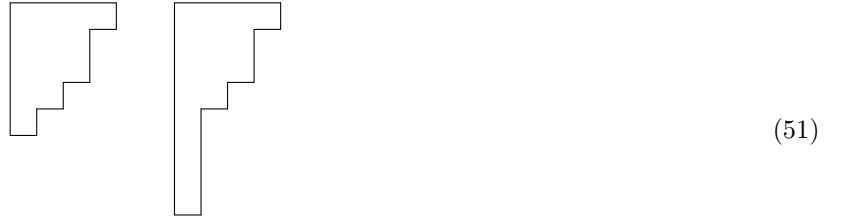
\begin{equation}\label{eq:lam-dias}
  \includegraphics{young19}\qquad
  \raisebox{-30bp}{\includegraphics{young20}}
\end{equation}
correspond to the pair of $\lo(2 k)$ Young diagrams
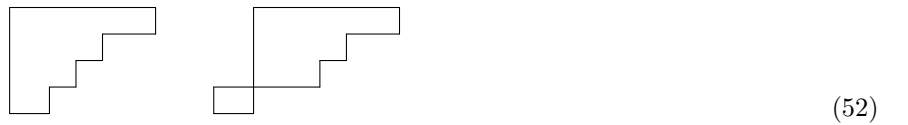
\begin{equation}
  \includegraphics{young21}\qquad\includegraphics{young22}
\end{equation}
in this order. To this end consider a self-non-complementary $\gO(d)$
Young diagram $\lambda$ with at most $k$ columns. Analogous to the
function $\chi^\lambda_\text{hw}$ defined in Sec.~\ref{sec:sl-hw-ten}
one can define a state
\begin{equation}\label{eq:phi_hw}
  \phi^\lambda_\text{hw}
    = \left( \prod_i a^\dagger_{\rho(i)\kappa(i)} \right)  |\rangle,
\end{equation}
where $\kappa(i)$ is the ordinal number of the column containing the
number $i$ in any fixed tableau assigned to $\lambda$. (Such states
are also considered in Ref.~\onlinecite{ref:Row11}.) By copying the
reasoning in Sec.~\ref{sec:O&o-ten} one finds that
$\phi^\lambda_\text{hw}$ is a highest-weight state of an $\lo(d)$
module belonging to the $\lo(d)$ irrep contained in the $\gO(d)$
irrep described by $\lambda$. It is easily calculated that
$r \phi^\lambda_\text{hw} = \phi^\lambda_\text{hw}$ for
$\tilde \lambda_1 < d/2$ and
$r \phi^\lambda_\text{hw} = - \phi^\lambda_\text{hw}$ for
$\tilde \lambda_1 > d/2$. Thus $\phi^\lambda_\text{hw}$ generates an
$\gO(d)$ module with Young diagram $\lambda$.

Now let the definitions of $a^\dagger_{p\tau}$ and $a_{p\tau}$ be
extended to negative $\tau$ by
\begin{equation}\label{eq:neg_tau}
  a^\dagger_{p\tau} = a_{p^*,-\tau} , \quad
  - k \le \tau \le k , \tau \ne 0 ,
\end{equation}
which gives the commutation relations
\begin{equation}
  \{ a^\dagger_{p\tau} , a_{q\upsilon} \} = \delta_{p\tau,q\upsilon} ,
\end{equation}
valid for every pair of $\tau$ and $\upsilon$ in the range of $\tau$
in Eq.~\eqref{eq:neg_tau}. It follows that when this range is ordered
from $-k$ to $k$ with 0 omitted, the operators
\begin{equation}\label{eq:f}
  f_{\tau\upsilon} 
    = \tfrac12 \sum_p [ a^\dagger_{p\tau} , a_{p\upsilon} ] ,
\end{equation}
obey the same commutation relations in terms of ordinal numbers as do
the transformations $\bar e_{pq}$ defined by Eq.~\eqref{eq:ebar} in
terms of the ordering of the range of $p$ from 1 to $d$. In particular
the operators $f_{\tau\upsilon} , \tau \le \upsilon ,$ span a Borel
subalgebra $\lb$ of $\lo(2 k)$. The derived Lie algebra $\lb'$ of
$\lb$ is spanned by the operators
$f_{\tau\upsilon} , \tau < \upsilon$. There are two kinds of these
operators
\begin{equation}
  \sum_p a_{p\tau} a^\dagger_{p\upsilon} , \quad
  \sum_p a_{p^*\tau} a_{p\upsilon} , \quad
  \tau > \upsilon > 0 .
\end{equation}
Acting on $\phi^\lambda_\text{hw}$, the operators of the first kind
attempt to move fermions in states $|p \tau \rangle$ into states with
the same $p$ and lower $\tau$, which is impossible because these
states are already occupied. The operators of the second kind attempt
to annihilate pairs of fermions in pairs of states $|p \tau \rangle$
and $|q \upsilon \rangle$ with $\tau \ne \upsilon$ and
$p + q = d + 1$, which is also impossible because
$\tilde \lambda_\tau + \tilde \lambda_\upsilon \le d$ for any pair of
different $\tau$ and $\upsilon$. In conclusion,
$\lb' \phi^\lambda_\text{hw} = 0$, and $\phi^\lambda_\text{hw}$ is
also an $\lo(2 k)$ highest-weight state. One can now calculate the
eigenvalues on $\phi^\lambda_\text{hw}$ of the basic operators
\begin{equation}
  f_{-\tau,-\tau} = \frac12 \sum_p [a^\dagger_{p,-\tau} , a_{p,-\tau}]
    = d/2 - \sum_p  a^\dagger_{p\tau} a_{p\tau} , \quad \tau > 0 ,
\end{equation}
of the Cartan subalgebra and arrive at the
relation~\eqref{eq:w-lambda}. It follows that $\phi^\lambda_\text{hw}$
generates an irreducible module over $\gO(d) \otimes \lo(2 k)$ whose
$\lambda$ and $w$ are combined according to Eq.~\eqref{eq:w-lambda}.
Since $\phi^\lambda_\text{hw}$ exists for every such pair of $\lambda$
and $w$, and, by Theorem~\ref{th:neer}, each $\lo(2 k)$ irrep with a
self-complementary Young diagram appears exactly once in combination
with a given $\lo(d)$ irrep, it follows that this $\lo(d)$ module must
be embedded in an $\gO(d)$ module belonging to the irrep given by
Eq.~\eqref{eq:w-lambda}. The entire $\gO(d)$--$\lo(2 k)$ case of
Theorem~\ref{th:howe} and the rule~\eqref{eq:w-lambda} are thus seen
to follow from Theorem~\ref{th:neer} in combination with the analysis
above of the action of a reflection.

When the $\gO(d)$ Young diagram $\lambda$ is self-complementary, and
$\lambda'$ is the Young diagram which, in the analysis of
Sec.~\ref{sec:O&o-ten}, produces the $\lo(d)$ highest-weight function
$\chi^{\lambda'}_\text{hw}$ with negative eigenvalue of
$\bar e_{d/2,d/2}$, then
$\phi^{\lambda'}_\text{hw} = \pm r \phi^\lambda_\text{hw}$, where the
sign depends on the Young tableaux. This confirms that the $\lo(d)$
modules generated by $\phi^\lambda_\text{hw}$ and
$\phi^{\lambda'}_\text{hw}$ combine to an $\gO(d)$ module.

\section{\label{sec:refl}Reflection of $\lo(2 k)$}

To see which transformation connects the states
$\phi^\lambda_\text{hw}$ and $\phi^{\lambda'}_\text{hw}$ corresponding
to the diagrams $\lambda$ and $\lambda'$ in Fig.~\eqref{eq:lam-dias},
consider the linear transformation of $\lA$ which acts distributively
on products and is generated in terms of a formal similarity map
$x \mapsto \sigma x \sigma^{-1}$ by
\begin{equation}\label{eq:sigma+}
  \sigma a^\dagger_{p1}  \sigma^{-1} = a_{p^*1} , \quad
  \sigma a_{p1} \sigma^{-1} = a^\dagger_{p^*1} , \quad
  \sigma a^\dagger_{p\tau} \sigma^{-1} = a^\dagger_{p\tau} , \quad
  \sigma a_{p\tau} \sigma^{-1} = a_{p\tau} , \quad
  \tau > 1 .
\end{equation}
This preserves the commutation relations~\eqref{eq:com}, and because
$(p^*)^* = p$, one gets $\sigma^2 a \sigma^{-2} = a$ for every
$a \in A$, so $x \mapsto \sigma x \sigma^{-1}$ is an involution of
$\lA$, that is, equal to its inverse map. Setting
\begin{equation}\label{eq:sigma}
  \sigma |\rangle
  = \left( \prod_p a^\dagger_{p1} \right) |\rangle
\end{equation}
and taking literally the formal similarity map, one obtains
\begin{equation}
  \sigma^2 |\rangle = \left( \prod_p a_{p^*1} \right)
\left( \prod_p a^\dagger_{p1} \right) |\rangle = |\rangle ,
\end{equation}
so $\sigma$ is an involution of $\Phi$. It is easily verified that
$x \mapsto \sigma x \sigma^{-1}$ also preserves the span of the
operators $f_{\tau\upsilon}$ given by Eq.~\eqref{eq:f}, which is
$\lo(2 k)$. Because it preserves commutation relations within $\lA$,
it preserves, in particular, the commutation relations in $\lo(2 k)$,
so it provides an involutionary automorphism of $\lo(2 k)$, which
may be called a reflection of $\lo(2 k)$.

The transformation
$a \mapsto r a r^{-1}$ maps $a^\dagger_{(d+1)/2,\tau}$ to
$-a^\dagger_{(d+1)/2,\tau}$ when $d$ is odd, and
$a^\dagger_{d/2,\tau}$ and $a^\dagger_{d/2+1,\tau}$ to one another
when $d$ is even, and does not change any other $a^\dagger_{p\tau}$,
and similarly for the annihilation operators. It follows that the
transformations $a \mapsto r a r^{-1}$ and
$a \mapsto \sigma a \sigma^{-1}$ commute for every $a \in A$. Further,
\begin{equation}
  \sigma r |\rangle = \sigma |\rangle
  = \left( \prod_p a^\dagger_{p1} \right) |\rangle , \quad
  r \sigma |\rangle
  = r \left( \prod_p a^\dagger_{p1} \right) |\rangle
  = - \left( \prod_p a^\dagger_{p1} \right) |\rangle ,
\end{equation}
so $\sigma r = - r \sigma$. Because, for even $d$, one gets
\begin{equation}
  \sigma \left( \prod_{p=1}^{d/2} a^\dagger_{p1} \right) |\rangle
  = \left( \prod_{p=1}^{d/2} a_{p^*1} \right)
    \left( \prod_p a^\dagger_{p1} \right) |\rangle
  = (-)^{d/2} \left( \prod_{p=1}^{d/2} a^\dagger_{p1} \right)
    |\rangle ,
\end{equation}
the highest-weight state $\phi^\lambda_\text{hw}$ corresponding to a
self-complementary $\gO(d)$ Young diagram $\lambda$ is an eigenstate
of $\sigma$ with eigenvalue $(-1)^{d/2}$. By
$\phi^{\lambda'}_\text{hw} = \pm r \phi^\lambda_\text{hw}$ and
$\sigma r = - r \sigma$, it follows that the corresponding state
$\phi^{\lambda'}_\text{hw}$ with a negative eigenvalue of
$\bar e_{d/2,d/2}$ is an eigenstate of $\sigma$ with eigenvalue
$(-1)^{d/2+1}$. The eigenvalue of $\sigma$ thus distinguishes the two
$\lo(d)$ irreps associated with a common $\lo(2 k)$ irrep from one
another in the same way as the eigenvalue of $r$ distinguishes the two
$\lo(2 k)$ irreps associated with a common $\lo(d)$ irrep from one
another. As $r$ connects the two former, and $\sigma$ the two latter,
a symmetry between the actions of $r$ and $\sigma$ with respect to
$\lo(d)$ and $\lo(2 k)$ is revealed. The transformation $\sigma$ may
be seen to be closely similar to the transformation with this symbol
employed by Weyl in his analysis of the restriction from $\gO(d)$ to
$\gSO(d)$.\cite{ref:Wey39}

The similarity of $\sigma$ to a reflection is displayed even more
clearly when one looks at the linear map $a \mapsto [x , a]$ with
$x \in \lo(2 k)$ and $a \in A$. It preserves each subspace $A_p$ of
$A$ spanned for a fixed $p$ by the operators
$a^\dagger_{p\tau} , -k \le \tau \le k , \tau \ne 0$, and its matrix
elements in the basis of these operators does not depend on $p$. The
same holds for the map $a \mapsto \sigma a \sigma^{-1}$. In a sense,
one could thus view our system as a system of $d$ fermion fields
living in a common space $U$ isomorphic to every $A_p$. The action of
$\sigma$ on $A_p$ results in an interchange of the basic operators
$a^\dagger_{p1}$ and $a^\dagger_{p,-1}$, which is seen to correspond
to the action of $r$ on $V$ according to Eq.~\eqref{eq:r} then $d$ is
even. The system of a single kind of fermion field living in $U$ in
this sense is the $d = 1$ case of the general system. For $d = 1$, the
Fock space $\Phi$ is isomorphic to the $2^k$-dimensional spinor space,
which carries a faithful representation of the double covering group
$\gPin(2 k)$ of $\gO(2 k)$.~\cite{ref:Bra35,ref:Goo98} Because
$\lo(2 k)$ is the Lie algebra of $\gPin(2 k)$, the symmetry of
Theorem~\ref{th:neer} with respect to $\lo(d)$ and $\lo(2 k)$ suggests
the existence of a $\lo(d)$--$\gPin(2 k)$ duality analogous to the
$\gO(d)$--$\lo(2 k)$ duality. Settling this matter would require an
analysis, which I shall not pursue, of the action relative
to the said realization of $\gPin(2 k)$ of $\sigma$ and the present
realization of $\lo(2 k)$. For $k = 1$, the transformation
$\sigma$ is similar to a particle-hole conjugation. Contrary to a
claim in Ref.~\onlinecite{ref:Nee19} it is different, however, from
the particle-hole conjugation $\gamma$ of
Refs.~\onlinecite{ref:Con35,ref:Rac42,ref:Bel59}, which obeys
$\gamma^2 a \gamma^{-2} = - a$ for $a \in A$ and only applies for even
$d$.\cite{ref:Nee15}

In Ref.~\onlinecite{ref:Nee19}, I employed a particular instance of
$\sigma$. It is instructive to review this example on the background
of the present, general definition. The system considered is the
atomic shell of Sec.~\ref{sec:GL-GL}. The variables $p$ and $\tau$ are
the magnetic quantum numbers $m_l$ and $m_s$, and $\sigma$ swaps
emptiness and occupation of 1-electron states with $m_s = -1/2$. The
corresponding map $x \mapsto \sigma x \sigma^{-1}$ is shown in
Ref.~\onlinecite{ref:Nee19} to transform the total spin $\bm S$ into a
``spin quasi-spin'' $\bm Q$. The Lie algebra $\lo(2 k) = \lo(4)$ is
spanned by the components of $\bm S$ and $\bm Q$, the components of
each of them span an $\lo(3)$ Lie algebra, and these $\lo(3)$ Lie
algebras commute. I call them $\lo(3)_S$ and $\lo(3)_Q$, and the row
lengths of their 1-row Young diagrams $S$ and $Q$. (The former is the
usual quantum number of total spin. The analogon of
$\lo(3)_S \simeq \lsl(2)$ for arbitrary $k$ is the $\lsl(k)$
subalgebra of the number conserving $\lgl(k)$ subalgebra of
$\lo(2 k)$. Only for $k = 2$ does a commuting and non-Abelian
subalgebra exist. For $k = 1$ one has $\lsl(k) = \lsl(1) = \{ 0 \}$,
and $\lo(2 k) = \lo(2)$ is 1-dimensional, and for $k \ge 3$ the Lie
algebra $\lo(2 k)$ is simple.)\cite{ref:Jac62} One gets
\begin{equation}
  Q_0 = \sigma \left( \sum_p m_{sp} \right) \sigma^{-1}
    = \tfrac12 ( n - d ) ,
\end{equation}
where $n$ is the number of electrons. Two other members of a basis for
$\lo(3)_Q$ raise or lower $n$ by 2 units. The row lengths of the Young
diagram of an $\lo(4)$ irrep are $w_{1,2} = Q \pm S$. This sheds light
on Racah's original definition of seniority,\cite{ref:Rac43} mentioned
in Sec.~\ref{sec:Sp-Sp}. The only operators in $\lo(4)$ which change
the number of electrons belong to $\lo(3)_Q$, so the sequence of
states with constant seniority $v$ according to Racah's definition has
constant $Q$. The leading state has
$\frac12 (v - d) = \frac12 (n - d) = Q_0 = -Q$, or $v = d - 2 Q$, so
$v$ and $Q$ are actually equivalent quantum numbers. One further gets
$v = d - w_1 - w_2$, which is the area of the $O(d)$ Young diagram
$\lambda$ in Fig.~\eqref{eq:rowe}. At the time of writing
Ref.~\onlinecite{ref:Nee19}, I was unaware of this and only saw that
$v$ generally differs from the area of the $\lo(d)$ diagram $\lambda$
in Fig.~\eqref{eq:neer}. This seemed to make this case different from
others such as that of the $\gSp(d)$--$\gSp(4)$ duality pertaining to
the systems of neutrons of protons in a nuclear $j$ shell, where an
appropriately defined seniority equals the area of the Young diagram
of the irrep of the number conserving group. In fact, in the the
atomic system, Racah's seniority $v$ is also the depth of the 1-column
Young diagram of the $\gSp(d)$ group arising in $jj$ coupling, so $Q$
equals Kerman's quasispin,\cite{ref:Ker61} as well.

\section{\label{sec:con}Concluding remarks}

The most important result of this study, in the view of the author, is
the demonstration in Sec.~\ref{sec:o-o&O-o} that Helmers's method of
calculation of characters\cite{ref:Hel61} provides a proof of the
$\gO(d)$--$\lo(2 k)$ special case of Theorem~\ref{th:howe} and an
explicit association of the participating irreps of $\gO(d)$ and
$\lo(2 k)$ when combined with an analysis of the representation of a
reflection. It is an open question whether this method could be
adapted to the boson case, where Weyl's character
formula\cite{ref:Wey25,ref:Wey26a,ref:Wey26b,
  ref:Fre54a,ref:Fre54b,ref:Fre56} is not available due to the
infinite dimensions of the irreps of the number non-conserving Lie
algebras. Also the Young diagrams introduced in Sec.~\ref{sec:spin} to
describe irreps of an orthogonal Lie algebra appear to be new in the
literature. The properties of the reflection $\sigma$ defined in
Sec.~\ref{sec:refl} further corroborate the picture,
already emerging from my study in Ref.~\onlinecite{ref:Nee19}, of
an almost perfect symmetry between $\lo(d)$ and $\lo(2 k)$ in their
dual relationship.

\begin{acknowledgments}

I am indebted to Roger Howe for providing me with copies of some of
his publications, which I could not access otherwise, including
Ref.~\onlinecite{ref:How95}. The data that support the findings of
this study are available within the article.

\end{acknowledgments}

\bibliography{note}

\begin{thebibliography}{38}%
\makeatletter
\providecommand \@ifxundefined [1]{%
 \@ifx{#1\undefined}
}%
\providecommand \@ifnum [1]{%
 \ifnum #1\expandafter \@firstoftwo
 \else \expandafter \@secondoftwo
 \fi
}%
\providecommand \@ifx [1]{%
 \ifx #1\expandafter \@firstoftwo
 \else \expandafter \@secondoftwo
 \fi
}%
\providecommand \natexlab [1]{#1}%
\providecommand \enquote  [1]{``#1''}%
\providecommand \bibnamefont  [1]{#1}%
\providecommand \bibfnamefont [1]{#1}%
\providecommand \citenamefont [1]{#1}%
\providecommand \href@noop [0]{\@secondoftwo}%
\providecommand \href [0]{\begingroup \@sanitize@url \@href}%
\providecommand \@href[1]{\@@startlink{#1}\@@href}%
\providecommand \@@href[1]{\endgroup#1\@@endlink}%
\providecommand \@sanitize@url [0]{\catcode `\\12\catcode `\$12\catcode
  `\&12\catcode `\#12\catcode `\^12\catcode `\_12\catcode `\%12\relax}%
\providecommand \@@startlink[1]{}%
\providecommand \@@endlink[0]{}%
\providecommand \url  [0]{\begingroup\@sanitize@url \@url }%
\providecommand \@url [1]{\endgroup\@href {#1}{\urlprefix }}%
\providecommand \urlprefix  [0]{URL }%
\providecommand \Eprint [0]{\href }%
\providecommand \doibase [0]{http://dx.doi.org/}%
\providecommand \selectlanguage [0]{\@gobble}%
\providecommand \bibinfo  [0]{\@secondoftwo}%
\providecommand \bibfield  [0]{\@secondoftwo}%
\providecommand \translation [1]{[#1]}%
\providecommand \BibitemOpen [0]{}%
\providecommand \bibitemStop [0]{}%
\providecommand \bibitemNoStop [0]{.\EOS\space}%
\providecommand \EOS [0]{\spacefactor3000\relax}%
\providecommand \BibitemShut  [1]{\csname bibitem#1\endcsname}%
\let\auto@bib@innerbib\@empty
\bibitem [{\citenamefont {Neerg{\aa}rd}(2019)}]{ref:Nee19}%
  \BibitemOpen
  \bibfield  {author} {\bibinfo {author} {\bibfnamefont {K.}~\bibnamefont
  {Neerg{\aa}rd}},\ }\href@noop {} {\bibfield  {journal} {\bibinfo  {journal}
  {J. Math. Phys.}\ }\textbf {\bibinfo {volume} {60}},\ \bibinfo {pages}
  {081705} (\bibinfo {year} {2019})}\BibitemShut {NoStop}%
\bibitem [{\citenamefont {Howe}(1976)}]{ref:How89}%
  \BibitemOpen
  \bibfield  {author} {\bibinfo {author} {\bibfnamefont {R.}~\bibnamefont
  {Howe}},\ }\href@noop {} {\bibfield  {journal} {\bibinfo  {journal} {Trans.
  Am. Math. Soc.}\ }\textbf {\bibinfo {volume} {313}},\ \bibinfo {pages} {539}
  (\bibinfo {year} {1989 (preprint 1976)})}\BibitemShut {NoStop}%
\bibitem [{\citenamefont {Helmers}(1961)}]{ref:Hel61}%
  \BibitemOpen
  \bibfield  {author} {\bibinfo {author} {\bibfnamefont {K.}~\bibnamefont
  {Helmers}},\ }\href@noop {} {\bibfield  {journal} {\bibinfo  {journal} {Nucl.
  Phys.}\ }\textbf {\bibinfo {volume} {23}},\ \bibinfo {pages} {594} (\bibinfo
  {year} {1961})}\BibitemShut {NoStop}%
\bibitem [{\citenamefont {Jacobsen}(1962)}]{ref:Jac62}%
  \BibitemOpen
  \bibfield  {author} {\bibinfo {author} {\bibfnamefont {N.}~\bibnamefont
  {Jacobsen}},\ }\href@noop {} {\emph {\bibinfo {title} {Lie Algebras}}}\
  (\bibinfo  {publisher} {Interscience Publishers},\ \bibinfo {address} {New
  York, USA},\ \bibinfo {year} {1962})\BibitemShut {NoStop}%
\bibitem [{\citenamefont {Weyl}(1939)}]{ref:Wey39}%
  \BibitemOpen
  \bibfield  {author} {\bibinfo {author} {\bibfnamefont {H.}~\bibnamefont
  {Weyl}},\ }\href@noop {} {\emph {\bibinfo {title} {The Classical Groups.
  Their Invariants and Representations}}}\ (\bibinfo  {publisher} {Princeton
  University Press},\ \bibinfo {address} {Princeton, USA},\ \bibinfo {year}
  {1939})\BibitemShut {NoStop}%
\bibitem [{\citenamefont {Howe}(1995)}]{ref:How95}%
  \BibitemOpen
  \bibfield  {author} {\bibinfo {author} {\bibfnamefont {R.}~\bibnamefont
  {Howe}},\ }\href@noop {} {\bibfield  {journal} {\bibinfo  {journal} {Isr.
  Math. Conf. Proc.}\ }\textbf {\bibinfo {volume} {8}},\ \bibinfo {pages} {1}
  (\bibinfo {year} {1995})}\BibitemShut {NoStop}%
\bibitem [{\citenamefont {Young}(1900)}]{ref:You00}%
  \BibitemOpen
  \bibfield  {author} {\bibinfo {author} {\bibfnamefont {A.}~\bibnamefont
  {Young}},\ }\href@noop {} {\bibfield  {journal} {\bibinfo  {journal} {Proc.
  London Math. Soc.}\ }\textbf {\bibinfo {volume} {s1-33}},\ \bibinfo {pages}
  {97} (\bibinfo {year} {1900})}\BibitemShut {NoStop}%
\bibitem [{\citenamefont {Young}(1902)}]{ref:You02}%
  \BibitemOpen
  \bibfield  {author} {\bibinfo {author} {\bibfnamefont {A.}~\bibnamefont
  {Young}},\ }\href@noop {} {\bibfield  {journal} {\bibinfo  {journal} {Proc.
  London Math. Soc.}\ }\textbf {\bibinfo {volume} {s1-34}},\ \bibinfo {pages}
  {361} (\bibinfo {year} {1902})}\BibitemShut {NoStop}%
\bibitem [{\citenamefont {Schur}(1901)}]{ref:Sch01}%
  \BibitemOpen
  \bibfield  {author} {\bibinfo {author} {\bibfnamefont {I.}~\bibnamefont
  {Schur}},\ }\href@noop {} {\emph {\bibinfo {title} {doctoral dissertation}}}\
  (\bibinfo {address} {Universit\"at Berlin, Germany},\ \bibinfo {year}
  {1901})\BibitemShut {NoStop}%
\bibitem [{\citenamefont {Weyl}(1928)}]{ref:Wey28}%
  \BibitemOpen
  \bibfield  {author} {\bibinfo {author} {\bibfnamefont {H.}~\bibnamefont
  {Weyl}},\ }\href@noop {} {\emph {\bibinfo {title} {Gruppentheorie und
  Quantenmechanik}}}\ (\bibinfo  {publisher} {Hirzel},\ \bibinfo {address}
  {Leipzig, Germany},\ \bibinfo {year} {1928})\BibitemShut {NoStop}%
\bibitem [{\citenamefont {Frobenius}(1903)}]{ref:Fro03}%
  \BibitemOpen
  \bibfield  {author} {\bibinfo {author} {\bibfnamefont {F.~G.}\ \bibnamefont
  {Frobenius}},\ }\href@noop {} {}\bibinfo {howpublished} {Sitzungsber.
  Preu\ss. Akad.} (\bibinfo {year} {1903}),\ \bibinfo {note} {328}\BibitemShut
  {NoStop}%
\bibitem [{\citenamefont {Frobenius}(1896)}]{ref:Fro96}%
  \BibitemOpen
  \bibfield  {author} {\bibinfo {author} {\bibfnamefont {F.~G.}\ \bibnamefont
  {Frobenius}},\ }\href@noop {} {}\bibinfo {howpublished} {Sitzungsber.
  Preu\ss. Akad.} (\bibinfo {year} {1896}),\ \bibinfo {note} {985}\BibitemShut
  {NoStop}%
\bibitem [{\citenamefont {Schur}(1905)}]{ref:Sch05}%
  \BibitemOpen
  \bibfield  {author} {\bibinfo {author} {\bibfnamefont {I.}~\bibnamefont
  {Schur}},\ }\href@noop {} {}\bibinfo {howpublished} {Sitzungsber. Preu\ss.
  Akad.} (\bibinfo {year} {1905}),\ \bibinfo {note} {406}\BibitemShut {NoStop}%
\bibitem [{\citenamefont {Rowe}, \citenamefont {Repka},\ and\ \citenamefont
  {Carvalho}(2011)}]{ref:Row11}%
  \BibitemOpen
  \bibfield  {author} {\bibinfo {author} {\bibfnamefont {D.~J.}\ \bibnamefont
  {Rowe}}, \bibinfo {author} {\bibfnamefont {J.}~\bibnamefont {Repka}}, \ and\
  \bibinfo {author} {\bibfnamefont {M.~J.}\ \bibnamefont {Carvalho}},\
  }\href@noop {} {\bibfield  {journal} {\bibinfo  {journal} {J. Math. Phys.}\
  }\textbf {\bibinfo {volume} {52}},\ \bibinfo {pages} {013507} (\bibinfo
  {year} {2011})}\BibitemShut {NoStop}%
\bibitem [{\citenamefont {Racah}(1943)}]{ref:Rac43}%
  \BibitemOpen
  \bibfield  {author} {\bibinfo {author} {\bibfnamefont {G.}~\bibnamefont
  {Racah}},\ }\href@noop {} {\bibfield  {journal} {\bibinfo  {journal} {Phys.
  Rev.}\ }\textbf {\bibinfo {volume} {63}},\ \bibinfo {pages} {367} (\bibinfo
  {year} {1943})}\BibitemShut {NoStop}%
\bibitem [{\citenamefont {Racah}(1949)}]{ref:Rac49}%
  \BibitemOpen
  \bibfield  {author} {\bibinfo {author} {\bibfnamefont {G.}~\bibnamefont
  {Racah}},\ }\href@noop {} {\bibfield  {journal} {\bibinfo  {journal} {Phys.
  Rev.}\ }\textbf {\bibinfo {volume} {76}},\ \bibinfo {pages} {1352} (\bibinfo
  {year} {1949})}\BibitemShut {NoStop}%
\bibitem [{\citenamefont {Kerman}(1961)}]{ref:Ker61}%
  \BibitemOpen
  \bibfield  {author} {\bibinfo {author} {\bibfnamefont {A.~K.}\ \bibnamefont
  {Kerman}},\ }\href@noop {} {\bibfield  {journal} {\bibinfo  {journal} {Ann.
  Phys. (N. Y.)}\ }\textbf {\bibinfo {volume} {12}},\ \bibinfo {pages} {300}
  (\bibinfo {year} {1961})}\BibitemShut {NoStop}%
\bibitem [{\citenamefont {Talmi}(1993)}]{ref:Tal93}%
  \BibitemOpen
  \bibfield  {author} {\bibinfo {author} {\bibfnamefont {I.}~\bibnamefont
  {Talmi}},\ }\href@noop {} {\emph {\bibinfo {title} {Simple Models of Complex
  Nuclei}}}\ (\bibinfo  {publisher} {Harwood Academic Publishers},\ \bibinfo
  {address} {Chur, Switzerland},\ \bibinfo {year} {1993})\BibitemShut {NoStop}%
\bibitem [{\citenamefont {Flowers}\ and\ \citenamefont
  {Szpikowski}(1964{\natexlab{a}})}]{ref:Flo64a}%
  \BibitemOpen
  \bibfield  {author} {\bibinfo {author} {\bibfnamefont {B.~H.}\ \bibnamefont
  {Flowers}}\ and\ \bibinfo {author} {\bibfnamefont {S.}~\bibnamefont
  {Szpikowski}},\ }\href@noop {} {\bibfield  {journal} {\bibinfo  {journal}
  {Proc. Phys. Soc.}\ }\textbf {\bibinfo {volume} {84}},\ \bibinfo {pages}
  {193} (\bibinfo {year} {1964}{\natexlab{a}})}\BibitemShut {NoStop}%
\bibitem [{\citenamefont {Flowers}\ and\ \citenamefont
  {Szpikowski}(1964{\natexlab{b}})}]{ref:Flo64b}%
  \BibitemOpen
  \bibfield  {author} {\bibinfo {author} {\bibfnamefont {B.~H.}\ \bibnamefont
  {Flowers}}\ and\ \bibinfo {author} {\bibfnamefont {S.}~\bibnamefont
  {Szpikowski}},\ }\href@noop {} {\bibfield  {journal} {\bibinfo  {journal}
  {Proc. Phys. Soc.}\ }\textbf {\bibinfo {volume} {84}},\ \bibinfo {pages}
  {673} (\bibinfo {year} {1964}{\natexlab{b}})}\BibitemShut {NoStop}%
\bibitem [{\citenamefont {Rowe}\ and\ \citenamefont
  {Carvalho}(2007)}]{ref:Row07}%
  \BibitemOpen
  \bibfield  {author} {\bibinfo {author} {\bibfnamefont {D.~J.}\ \bibnamefont
  {Rowe}}\ and\ \bibinfo {author} {\bibfnamefont {M.~J.}\ \bibnamefont
  {Carvalho}},\ }\href@noop {} {\bibfield  {journal} {\bibinfo  {journal} {J.
  Phys. A; Math. Theor.}\ }\textbf {\bibinfo {volume} {40}},\ \bibinfo {pages}
  {471} (\bibinfo {year} {2007})}\BibitemShut {NoStop}%
\bibitem [{\citenamefont {Shale}(1962)}]{ref:Sha62}%
  \BibitemOpen
  \bibfield  {author} {\bibinfo {author} {\bibfnamefont {D.}~\bibnamefont
  {Shale}},\ }\href@noop {} {\bibfield  {journal} {\bibinfo  {journal} {Trans.
  Am. Math. Soc.}\ }\textbf {\bibinfo {volume} {103}},\ \bibinfo {pages} {149}
  (\bibinfo {year} {1962})}\BibitemShut {NoStop}%
\bibitem [{\citenamefont {Rosensteel}\ and\ \citenamefont
  {Rowe}(1977)}]{ref:Ros77}%
  \BibitemOpen
  \bibfield  {author} {\bibinfo {author} {\bibfnamefont {G.}~\bibnamefont
  {Rosensteel}}\ and\ \bibinfo {author} {\bibfnamefont {D.~J.}\ \bibnamefont
  {Rowe}},\ }\href@noop {} {\bibfield  {journal} {\bibinfo  {journal} {Phys.
  Rev. Lett.}\ }\textbf {\bibinfo {volume} {38}},\ \bibinfo {pages} {10}
  (\bibinfo {year} {1977})}\BibitemShut {NoStop}%
\bibitem [{\citenamefont {Iachello}\ and\ \citenamefont
  {Arima}(1980)}]{ref:Iac80}%
  \BibitemOpen
  \bibfield  {author} {\bibinfo {author} {\bibfnamefont {F.}~\bibnamefont
  {Iachello}}\ and\ \bibinfo {author} {\bibfnamefont {A.}~\bibnamefont
  {Arima}},\ }\href@noop {} {\emph {\bibinfo {title} {The Interacting Boson
  Model}}}\ (\bibinfo  {publisher} {Cambridge University Press},\ \bibinfo
  {address} {Cambridge, United Kingdom},\ \bibinfo {year} {1980})\BibitemShut
  {NoStop}%
\bibitem [{\citenamefont {{Lerma H.}}\ \emph {et~al.}(2006)\citenamefont
  {{Lerma H.}}, \citenamefont {Errea}, \citenamefont {Dukelsky}, \citenamefont
  {Pittel},\ and\ \citenamefont {Isacker}}]{ref:Ler06}%
  \BibitemOpen
  \bibfield  {author} {\bibinfo {author} {\bibfnamefont {S.}~\bibnamefont
  {{Lerma H.}}}, \bibinfo {author} {\bibfnamefont {B.}~\bibnamefont {Errea}},
  \bibinfo {author} {\bibfnamefont {J.}~\bibnamefont {Dukelsky}}, \bibinfo
  {author} {\bibfnamefont {S.}~\bibnamefont {Pittel}}, \ and\ \bibinfo {author}
  {\bibfnamefont {P.~V.}\ \bibnamefont {Isacker}},\ }\href@noop {} {\bibfield
  {journal} {\bibinfo  {journal} {Phys. Rev. C}\ }\textbf {\bibinfo {volume}
  {74}},\ \bibinfo {pages} {024314} (\bibinfo {year} {2006})}\BibitemShut
  {NoStop}%
\bibitem [{\citenamefont {Goodman}\ and\ \citenamefont
  {Wallach}(1998)}]{ref:Goo98}%
  \BibitemOpen
  \bibfield  {author} {\bibinfo {author} {\bibfnamefont {R.}~\bibnamefont
  {Goodman}}\ and\ \bibinfo {author} {\bibfnamefont {N.~R.}\ \bibnamefont
  {Wallach}},\ }\href@noop {} {\emph {\bibinfo {title} {Representations and
  Invariants of the Classical groups}}}\ (\bibinfo  {publisher} {Cambridge
  University Press},\ \bibinfo {address} {Cambridge, United Kingdom},\ \bibinfo
  {year} {1998})\BibitemShut {NoStop}%
\bibitem [{\citenamefont {Cartan}(1913)}]{ref:Car13}%
  \BibitemOpen
  \bibfield  {author} {\bibinfo {author} {\bibfnamefont {E.}~\bibnamefont
  {Cartan}},\ }\href@noop {} {\bibfield  {journal} {\bibinfo  {journal} {Bull.
  Soc. Math. France}\ }\textbf {\bibinfo {volume} {41}},\ \bibinfo {pages} {53}
  (\bibinfo {year} {1913})}\BibitemShut {NoStop}%
\bibitem [{\citenamefont {Brauer}\ and\ \citenamefont
  {Weyl}(1935)}]{ref:Bra35}%
  \BibitemOpen
  \bibfield  {author} {\bibinfo {author} {\bibfnamefont {R.}~\bibnamefont
  {Brauer}}\ and\ \bibinfo {author} {\bibfnamefont {H.}~\bibnamefont {Weyl}},\
  }\href@noop {} {\bibfield  {journal} {\bibinfo  {journal} {Am. J. Math.}\
  }\textbf {\bibinfo {volume} {57}},\ \bibinfo {pages} {425} (\bibinfo {year}
  {1935})}\BibitemShut {NoStop}%
\bibitem [{\citenamefont {Condon}\ and\ \citenamefont
  {Shortley}(1935)}]{ref:Con35}%
  \BibitemOpen
  \bibfield  {author} {\bibinfo {author} {\bibfnamefont {E.~U.}\ \bibnamefont
  {Condon}}\ and\ \bibinfo {author} {\bibfnamefont {G.~H.}\ \bibnamefont
  {Shortley}},\ }\href@noop {} {\emph {\bibinfo {title} {The Theory of Atomic
  Spectra}}}\ (\bibinfo  {publisher} {Cambridge University Press},\ \bibinfo
  {address} {Cambridge, United Kingdom},\ \bibinfo {year} {1935})\BibitemShut
  {NoStop}%
\bibitem [{\citenamefont {Racah}(1942)}]{ref:Rac42}%
  \BibitemOpen
  \bibfield  {author} {\bibinfo {author} {\bibfnamefont {G.}~\bibnamefont
  {Racah}},\ }\href@noop {} {\bibfield  {journal} {\bibinfo  {journal} {Phys.
  Rev.}\ }\textbf {\bibinfo {volume} {62}},\ \bibinfo {pages} {438} (\bibinfo
  {year} {1942})}\BibitemShut {NoStop}%
\bibitem [{\citenamefont {Bell}(1959)}]{ref:Bel59}%
  \BibitemOpen
  \bibfield  {author} {\bibinfo {author} {\bibfnamefont {J.~S.}\ \bibnamefont
  {Bell}},\ }\href@noop {} {\bibfield  {journal} {\bibinfo  {journal} {Nucl.
  Phys.}\ }\textbf {\bibinfo {volume} {12}},\ \bibinfo {pages} {117} (\bibinfo
  {year} {1959})}\BibitemShut {NoStop}%
\bibitem [{\citenamefont {Neerg{\aa}rd}(2015)}]{ref:Nee15}%
  \BibitemOpen
  \bibfield  {author} {\bibinfo {author} {\bibfnamefont {K.}~\bibnamefont
  {Neerg{\aa}rd}},\ }\href@noop {} {\bibfield  {journal} {\bibinfo  {journal}
  {Phys. Rev. C}\ }\textbf {\bibinfo {volume} {91}},\ \bibinfo {pages} {144313}
  (\bibinfo {year} {2015})}\BibitemShut {NoStop}%
\bibitem [{\citenamefont {Weyl}(1925)}]{ref:Wey25}%
  \BibitemOpen
  \bibfield  {author} {\bibinfo {author} {\bibfnamefont {H.}~\bibnamefont
  {Weyl}},\ }\href@noop {} {\bibfield  {journal} {\bibinfo  {journal} {Math.
  Z.}\ }\textbf {\bibinfo {volume} {23}},\ \bibinfo {pages} {271} (\bibinfo
  {year} {1925})}\BibitemShut {NoStop}%
\bibitem [{\citenamefont {Weyl}(1926{\natexlab{a}})}]{ref:Wey26a}%
  \BibitemOpen
  \bibfield  {author} {\bibinfo {author} {\bibfnamefont {H.}~\bibnamefont
  {Weyl}},\ }\href@noop {} {\bibfield  {journal} {\bibinfo  {journal} {Math.
  Z.}\ }\textbf {\bibinfo {volume} {24}},\ \bibinfo {pages} {328} (\bibinfo
  {year} {1926}{\natexlab{a}})}\BibitemShut {NoStop}%
\bibitem [{\citenamefont {Weyl}(1926{\natexlab{b}})}]{ref:Wey26b}%
  \BibitemOpen
  \bibfield  {author} {\bibinfo {author} {\bibfnamefont {H.}~\bibnamefont
  {Weyl}},\ }\href@noop {} {\bibfield  {journal} {\bibinfo  {journal} {Math.
  Z.}\ }\textbf {\bibinfo {volume} {24}},\ \bibinfo {pages} {377} (\bibinfo
  {year} {1926}{\natexlab{b}})}\BibitemShut {NoStop}%
\bibitem [{\citenamefont {Freudenthal}(1954{\natexlab{a}})}]{ref:Fre54a}%
  \BibitemOpen
  \bibfield  {author} {\bibinfo {author} {\bibfnamefont {H.}~\bibnamefont
  {Freudenthal}},\ }\href@noop {} {\bibfield  {journal} {\bibinfo  {journal}
  {Indag. Math.}\ }\textbf {\bibinfo {volume} {16}},\ \bibinfo {pages} {369}
  (\bibinfo {year} {1954}{\natexlab{a}})}\BibitemShut {NoStop}%
\bibitem [{\citenamefont {Freudenthal}(1954{\natexlab{b}})}]{ref:Fre54b}%
  \BibitemOpen
  \bibfield  {author} {\bibinfo {author} {\bibfnamefont {H.}~\bibnamefont
  {Freudenthal}},\ }\href@noop {} {\bibfield  {journal} {\bibinfo  {journal}
  {Indag. Math.}\ }\textbf {\bibinfo {volume} {16}},\ \bibinfo {pages} {487}
  (\bibinfo {year} {1954}{\natexlab{b}})}\BibitemShut {NoStop}%
\bibitem [{\citenamefont {Freudenthal}(1956)}]{ref:Fre56}%
  \BibitemOpen
  \bibfield  {author} {\bibinfo {author} {\bibfnamefont {H.}~\bibnamefont
  {Freudenthal}},\ }\href@noop {} {\bibfield  {journal} {\bibinfo  {journal}
  {Indag. Math.}\ }\textbf {\bibinfo {volume} {18}},\ \bibinfo {pages} {511}
  (\bibinfo {year} {1956})}\BibitemShut {NoStop}%
\end{thebibliography}%

\end{document}